\documentclass[a4paper,12pt]{article}
\usepackage{xcolor}
\usepackage{mathtools,amssymb}
\usepackage{verbatim}
\usepackage{comment}

\usepackage[numbers,sort&compress]{natbib}
\usepackage[colorlinks=true
  ,urlcolor=blue
  ,anchorcolor=blue
  ,citecolor=blue
  ,filecolor=blue
  ,linkcolor=blue
  ,menucolor=blue
  ,linktocpage=true
  ,pdfproducer=medialab
  ,pdfa=true
]{hyperref} 
\usepackage{uri}
\usepackage{tikz-cd}
\usepackage{graphicx}
\tikzcdset{scale cd/.style={every label/.append style={scale=#1},
    cells={nodes={scale=#1}}}}
\tikzset{every picture/.style={line width=0.75pt}} 

\usepackage[normalem]{ulem}

\usepackage{enumitem}
\usepackage{geometry}
\usepackage{jheppub}

\newcommand{\eq}[1]{\begin{align}#1\end{align}}

\newcommand{\tr}{\operatorname{tr}}

\numberwithin{equation}{section}

\title{Modular Hamiltonian and entanglement 
 entropy in the BMS free fermion theory}

\author[a,c]{Peng-Xiang Hao,}
\author[e,a]{Wen-Xin Lai,}
\author[a,b]{Wei Song,}
\author[a,d]{Zehua Xiao}

\emailAdd{pxhao@yukawa.kyoto-u.ac.jp}
\emailAdd{laiwenxin@ucas.ac.cn}
\emailAdd{wsong2014@mail.tsinghua.edu.cn}
\emailAdd{xiaozehua@cqcet.edu.cn}

\affiliation[a]{Yau Mathematical Sciences Center, Tsinghua University, Beijing 100084, China}
\affiliation[b]{Department of Mathematical Sciences, Tsinghua University, Beijing 100084, China}
\affiliation[c]{Center for Gravitational Physics and Quantum Information, Yukawa Institute for Theoretical Physics, Kyoto University, Kitashirakawa Oiwakecho, Sakyo-ku, Kyoto 606-8502, Japan}
\affiliation[d]{Chongqing Polytechnic University of Electronic Technology, Chongqing 401331, China}
\affiliation[e]{Kavli Institute for Theoretical Sciences, University of Chinese Academy of Sciences, Beijing 100190, China}

\abstract{%
We study the modular Hamiltonian and the entanglement entropy of the BMS-invariant free fermion model. 
Starting from the modular Hamiltonian on a half-line interval, we calculate the modular Hamiltonian for a region consisting of two disjoint intervals using the uniformization map and a background field method. The resulting modular Hamiltonian contains a local term which generates a geometrical flow, and a bi-local term which mixes operators between the two intervals. We further derive a relation between Wightman functions and the modular Hamiltonian using both diagonalization and a coherent state method. This enables us to compute the entanglement entropy for multi-disjoint intervals. 
Our explicit results in the BMS free fermion model are compatible with earlier ones based on symmetry and holography.
}

\keywords{}

\begin{document}
\maketitle

\newpage

\section{Introduction}

Recently, Carrollian conformal field theories (CCFTs) have become a playground for many inter-disciplinary studies, including quantum field theory, holographic duality, condensed matter,  representation theory, and quantum information. 
Just as Lorentzian invariant quantum field theories are formulated on Riemannian manifolds, Carrollian
invariant field theories \cite{levy1965nouvelle,sen1966analogue,Duval:2014uva,Duval:2014uoa,Duval:2014lpa} 
can be formulated on a Newton-Cartan manifold featuring a degenerate metric and a vector field along which a Carrollian boost symmetry exists. Such theories explore physics in the ultra-relativistic limit and are complementary to Galilean-invariant theories, which describe systems in the non-relativistic limit. 
From a purely field-theoretical perspective, both the Carrollian and Galilean field theories are relevant in condensed matter physics. For instance, Galilean symmetry can be found in cold atomic gas \cite{Bloch:2008zzb} and non-relativistic fluid systems \cite{Zhou:2016cnl}.
On the other hand, ultra-slow speed of light as low as 17~m/s can be observed in laser-cooled, ultra-cold atomic systems \cite{Marangos:1999,Hau:1999}, and lately Carrollian CFTs are also found to be related to the fracton phase of matter \cite{Figueroa-OFarrill:2023vbj,
Bidussi:2021nmp,Bagchi:2022eui,Figueroa-OFarrill:2023qty}. 
If a Carrollian theory is further invariant under scaling symmetries, we obtain Carrollian conformal field theories (CCFT). 
Carrollian conformal symmetry is isomorphic to the BMS symmetry --- the asymptotic symmetry \cite{Bondi:1962px,Sachs:1962wk,Barnich:2006av} of flat spacetime and thus plays an important role in understanding flat holography \cite{Barnich:2010eb,Bagchi:2010zz,Fareghbal:2013ifa,Cotler:2024cia}. By integrating out the null direction \cite{Donnay:2022wvx,Donnay:2022aba,Bagchi:2022emh,Bagchi:2023fbj}, the Carrollian approach to flat holography is connected to the celestial holography approach \cite{Pasterski:2016qvg,Pasterski:2017kqt}.

Discussions of CCFTs fall into two categories: i) focusing on general properties based on symmetry structure, or ii) delving into explicit models. 
The first approach is especially fruitful in two dimensions, where the Carrollian conformal group is infinite-dimensional. Similar to a CFT$_2$, Carrollian conformal symmetry can be used to discuss representations \cite{Bagchi:2009ca,Bagchi:2009pe}, correlators \cite{Bagchi:2010vw}, partition functions \cite{Oblak:2015sea,Poulias:2025eck}, BMS$_3$ blocks \cite{Bagchi:2016geg,Bagchi:2017cpu,ammon2021semi}, entanglement entropy \cite{Bagchi:2014iea,Jiang:2017ecm,  Hijano:2017eii}, etc. While the results obtained based on symmetry are universal, it is still necessary to understand how explicit models realize the symmetries and how the quantum theory behaves. One example is the representation of the BMS group. Purely based on the symmetries, it is most natural to consider the induced representation and the singlet highest weight representation as discussed in \cite{Bagchi:2009ca,Barnich:2014kra,Barnich:2015uva}. Further analyzing the symmetry algebra, it was pointed out that there exists a multiplet version of the highest weight representation for quasi-primary states \cite{Chen:2020vvn,Chen:2022jhx}.
In the BMS-invariant free boson and fermion theories \cite{Hao:2021urq,Hao:2022xhq}, however,  it was found that highest weight multiplets also appear in primary states, and they form a novel representation structure, the staggered BMS module. 
So far, the study of explicit models is still at an early stage. 
A Liouville-like theory featuring BMS symmetries has been developed in \cite{Barnich:2012rz,Barnich:2013yka,Barnich:2017jgw}. 
Free bosons and free fermion model have been constructed in \cite{Bagchi:2015nca,Bagchi:2017cte}, and the quantization has been carried out in \cite{Hao:2021urq,Chen:2021xkw,Bagchi:2022eav,Yu:2022bcp,Hao:2022xhq,Banerjee:2022ocj,deBoer:2023fnj,Cotler:2024xhb}.

In this paper, we study some important quantities in explicit models: entanglement entropy and modular Hamiltonian in BMS free fermions models \cite{Hao:2022xhq}.
Entanglement entropy is a useful measure of quantum entanglement between a subsystem and its complement. Based on symmetries,  entanglement entropy for a single interval in two-dimensional CCFT was calculated using the correlation function of twist operators \cite{Bagchi:2014iea} and the generalized Rindler method \cite{Jiang:2017ecm}. It was proposed in \cite{Jiang:2017ecm,Apolo:2020bld} that holographic entanglement entropy in flat holography can be obtained from the swing surface prescription. 
Modular Hamiltonian \cite{Apolo:2020qjm}, bulk reconstruction \cite{Hijano:2019qmi,Nguyen:2023vfz,Bagchi:2023cen,Chen:2023naw,Hao:2025btl}, entanglement wedge, reflected entropy and holographic negativity \cite{Basu:2021awn,Setare:2022emp,Liu:2023djf} have also been studied.

In this paper, we use the method of uniformization map and background field~\cite{Wong:2018svs} to derive the modular Hamiltonian for a subregion with multiple intervals $V_i$. The result is given by
\begin{align}
    \mathcal H&=i\int_{V} \int_{V} dxdx' \Big[\frac{\delta(\log z(x)-\log z(x'))}{2(x-x')}(\psi_1(x)\psi_2(x',y')+\psi_2(x,y)\psi_1(x'))\nonumber\\
    &+(y'-y)\,\partial_{x'}\left(\frac{\delta(\log z(x)-\log z(x'))}{(x-x')}\right)\psi_1(x)\psi_1(x')\nonumber\\
    &+\frac{k_+\delta( \frac{1}{z(x)}- \frac{1}{z(x')})-k_-\delta( z(x)- z(x'))}{x-x'}\psi_1(x)\psi_1(x')\Big],
\end{align}
where $x,x'\in V=\cup_{i} V_i$, and $z(x)$ is the uniformazation map 
\begin{align} z(x)\equiv -{\prod^n_{i=1}(x-a_i)\over \prod^n_{i=1}(x-b_i)}\,.\end{align}
Furthermore, we find the relation between the entanglement entropy and the Wightman function is given by \begin{align}
    \label{SEECintro}
    S_{\mathcal A}&=-\tr[(1-C)\log(1-C)+C\log C],
\end{align}
where $C$ is the Wightman function between the fermions.
In the highest weight vacuum, the entanglement entropy is given by
\begin{align}
    S_{\mathcal A}=&\frac{1}{6}\Big(\sum_{i,j}\log|b_i-a_j|-\sum_{i<j}\log|a_i-a_j|-\sum_{i<j}\log|b_i-b_j|-n\log\epsilon\Big).\label{Seef}
\end{align}
These results are similar to the results of free fermion CFT$_2$, and are also consistent with the swing surface proposal \cite{Apolo:2020bld}.

The layout of the rest of this paper is as follows. In section 2, we review the basic properties of CCFT and the BMS invariant free fermions model. In section 3, we derive the modular Hamiltonian for multi-intervals and discuss the modular flow. In section 4, we compute the entanglement entropy.

\section{The setup}

In this section, we provide a brief overview of certain aspects of 2D Galilean and Carrollian Conformal Field Theories (G/C CFTs), which are quantum field theories exhibiting Galilean or Carrollian conformal symmetries. These theories can be examined either intrinsically or as specific limits of two-dimensional conformal field theories.
The Galilean CFT is derived from the relativistic CFT by taking a non-relativistic limit, where the speed of light $c\rightarrow \infty$. 
Conversely, the Carrollian CFT, also known as Bondi-(van der Burg)-Metzner-Sachs field theory (BMSFT), is constructed from the relativistic CFT by taking an ultra-relativistic limit, where the speed of light $c\rightarrow 0$.  
In this paper, we will mainly study Galilean/Carrollian conformal field theory intrinsically, without referring to the NR/UR limit.

\subsection{The Galilean/Carrollian conformal symmetry}\label{Thesymmetry}
Relativistic quantum field theory can be constructed by covariantly coupling matter fields on a non-dynamical Riemannian geometry with local Lorentzian symmetry.
Similarly, quantum field theories in the non-relativistic or ultra-relativistic limit can be constructed on Newton-Cartan geometry or Carrollian geometry, both of which can be formulated within the framework of Cartan geometry \cite{AIHPA_1965__3_1_1_0, Duval:2014uoa,Hehl:2007bn,cartan1986manifolds}. 
The flat version of these geometries in two dimensions has translational symmetries and a boost symmetry, the latter of which leaves one direction invariant. Let us denote the boost invariant direction $x^1$. Then the boost transformation acts on the coordinates as
 \begin{gather} x^{\mu}\to \Lambda^{\mu}_{\ \nu}x^{\nu}, \quad \mu=1,2 \label{boost}\\
     \Lambda_{\ \nu}^{\mu}=\left[\,\begin{array}{cc}
1 & 0\\
v & 1
\end{array}\,\right]\nonumber
\end{gather} 
The $x^{1}$ coordinate is defined as the absolute time for the Galilean boost, while it is the absolute spatial direction for the Carrollian boost. 
Vectors are denoted with upper indices $V^{\mu}$, and transform under the same rule as \eqref{boost}. 
1-forms are denoted with lower indices and transform as $V_{\mu}\to V_{\nu}\Lambda^{\nu}_{\ \mu}$. 
Indices can be raised and lowered by the anti-symmetric tensor $\epsilon_{\mu\nu}$ and its inverse, so that $V_{\mu}=\epsilon_{\mu\nu} V^\nu$.
Higher rank tensors transform covariantly under translations and boosts. The basic geometric quantities of 2D Newton-Cartan geometry are a boost invariant 1-form $e_{\mu}=(1,0)$ and a boost invariant vector $\tau^{\mu}=(0,1)^{{T}}$, and the anti-symmetric tensor $\epsilon_{\mu\nu}$.

If we further assume scaling symmetry ${x}^{\mu}\to\lambda x^{\mu}$ in addition to the translational and boost symmetries in two dimensions, it has been shown that the theory also possesses Galilean conformal symmetry if the boost invariant direction $x^1$ is absolute time, or Carrollian conformal symmetry \cite{Chen:2019hbj,Chen:2020vvn,Hao:2022xhq} if $x^1$ is absolute space. The G(Galilean)  and C(Carrollian) conformal algebras are isomorphic to each other, and also to the BMS$_3$ algebra, which is the asymptotic symmetry algebra for flat spacetime in three dimensions. We will refer to the quantum field theories with these symmetries as G/C CFT henceforth.  We will also use the terminologies CCFT and BMSFT interchangeably. The G/C conformal transformation, which is also referred to as the BMS transformation, is given by
\cite{Barnich:2012xq}
\begin{align}
\tilde{x} = f(x),\quad \tilde{y} = f'(x)\,y+g(x),\label{bmstran001} 
\end{align}
where we have renamed the coordinates as  $x^1=x,\, x^2=y$ for convenience. 
The Noether currents generating the translational symmetries along $x^\nu$ are given by the stress tensor $T^\mu\,_{\nu}$ which takes the following form \cite{Chen:2019hbj, Bagchi:2021gai,Baiguera:2022lsw}  \begin{align}\label{Tuv}
T^{\mu}_{\ \ \nu}=&\frac{\tau^{\mu}}{2e}\frac{\delta S}{\delta\tau^{\nu}}+\frac{e^{\mu}}{2e}\frac{\delta S}{\delta e^{\nu}}\\
=&\left[\begin{array}{cc}
\frac{1}{2}M(x,y) & 0\\
-\frac{1}{2}T(x,y) & -\frac{1}{2}M(x,y)
\end{array}\right]    
\end{align}
and satisfies the conservation law,
\begin{equation} 
\partial_\mu  T^\mu_{\ \ \nu}=0, \quad \mu,\nu=1,2.
\label{currents}
\end{equation}
Using the conservation law, we find that translational invariance in $y$ implies that $M(x,y)$ is independent of $y$, and translational invariance of $x$ implies that  the  combination  
\begin{equation}
  L(x)=T(x,y)-y\partial_{x}M(x,y)  
\end{equation}
 is also $y$-independent. 
We will use $L$ and $M$ as the stress tensor henceforth.
Under the G/C conformal transformation \eqref{bmstran001} the BMS currents transform as follows:
\begin{align}\label{transf}
    \tilde{M}(\tilde x) =& \left({f}'\right)^{-2} \Big(M\left(x \right)-\dfrac{c_{M}}{12}\{f, x\}\Big),\\
    \tilde{L}(\tilde{x})  =&\left(f'\right)^{-2}\left(L(x)-\frac{c_{L}}{12}\left\{ f,x\right\} \right)-\left(f'\right)^{-2}\left\{ \left(\partial_{x}\frac{g}{f'}\right)M(x)+\partial_{x}\left[\frac{g}{f'}M(x)+\frac{c_{M}}{12}\partial_{x}^{2}\frac{g}{f'}\right]\right\} \nonumber
\end{align}
where $\{\tilde{f}, \tilde{x}\}$ is the Schwarzian derivative. The mode expansions of the currents $L(x),M(x)$ lead to an infinite number of symmetry generators $L_{n},M_{n},n\in\mathbb{Z}$, which form the following algebra:
\begin{align}
    [L_{n},L_{m}]&=(n-m)L_{n+m}+\frac{c_{L}}{12}n(n^{2}-1)\delta_{m+n,0},\nonumber\\
    [L_{n},M_{m}]&=(n-m)M_{n+m}+\frac{c_{M}}{12}n(n^{2}-1)\delta_{m+n,0},\label{bms}\\
    [M_{n},M_{m}]&=0.\nonumber
\end{align}
The above algebra is the two dimensional Carrollian conformal algebra $\widehat{\mathfrak{cca}}_{2}\cong\widehat{\mathfrak{bms}}_{3}$ if $x^1=x$ is spatial,  and Galilean conformal algebra $\widehat{\mathfrak{gca}}_{2} $ if $x$ is temporal.
Alternativaly, $\widehat{\mathfrak{cca}}_{2}$ and  $\widehat{\mathfrak{gca}}_{2}$ can also be
derived from the 2D Virasoro algebra by either taking the ultra-relativistic limit or the non-relativistic limit \cite{Bagchi:2023dzx}, respectively. In analogous to ${\mathfrak{sl}}2\oplus {\mathfrak{sl}}2 $ algebra in CFT$_2$, the sub-algebra  $L_i, M_i,i=0,\pm1$ is referred to as the global G/C conformal algebra. This subalgebra is also isomorphic to the Poincar\'e algebra in three dimensions.
Note that the generator $L_{0}$ is the momentum generator and $M_{0}$ is the Hamiltonian in  Carrollian CFTs, and the other way around in Galilean CFTs.

Based on the symmetries, entanglement entropy on a single interval in the highest vacuum of Carrollian conformal field theory has been discussed in \cite{Bagchi:2014iea, Jiang:2017ecm}, and holographically in \cite{Jiang:2017ecm}. 
The entanglement entropy of an interval $\mathcal{A}$ with end points $(a, u_a)$ and $(b, u_b)$ is given by
\begin{align}
    S_{E}=\frac{c_{L}}{6}\log\frac{b-a}{\epsilon}+\frac{c_{M}}{6}\left(\frac{u_b-u_a}{b-a}-\frac{\epsilon_{y}}{\epsilon}\right),\label{singleEE}
\end{align}
where $b-a>0,\, u_b-u_a>0$ describe the length scale of the subregion $\mathcal{A}$ and $\epsilon,\epsilon_{y}$ are the regulators in the two directions, respectively. 

The modular Hamiltonian \cite{Apolo:2020qjm} has also been written down based on the Rindler method \cite{Casini:2009vk, Castro:2015csg, Jiang:2017ecm}. With the same parameterization, the modular Hamiltonian is given by 
\begin{align}
    \mathcal{H} =-\int_{a}^{b}dx\left[X(x)L(x)+Y(x)M(x)\right]+const,
\end{align}
where the $L(x),M(x)$ are the aforementioned conserved currents, and the functions $X(x),Y(x)$ are given by 
\begin{align}\label{RindlerH}
X(x) & =\frac{\left(x-a\right)\left(x-b\right)}{a-b},\quad
Y(x) =\frac{\left[u_{b}\left(x-a\right)^{2}-u_{a}\left(x-b\right)^{2}\right]}{\left(b-a\right)^{2}}.
\end{align}

\subsection{The free fermion model}
Consider the 2D Galilean/Carrollian spinor group $\text{CSpin}(2)$. The Clifford algebra is given by
\begin{align}
    \{\Gamma^{\mu},\Gamma^{\nu}\}=2\tau^{\mu}\tau^{\nu}I,\label{gammaeq}
\end{align}
where $\tau^\mu=(0,1)^T$ is the boost invariant vector, and $\Gamma^{\mu}$ is the matrix representation of the Clifford algebra generators. As discussed in \cite{Hao:2022xhq},
there are two inequivalent classes of solutions to \eqref{gammaeq},  one of which leads to chiral fermion models and the other to the so-called BMS fermion models. 
We focus on the latter case in this paper, with the choice of gamma matrices
\begin{align}
    \Gamma^{1}=\left[\begin{array}{cc}
0 & 0\\
2 & 0
\end{array}\right],\quad\Gamma^{2}=\left[\begin{array}{cc}
1 & 0\\
0 & -1
\end{array}\right].
\end{align}
The 2D G/C spinor field has two components $\psi=(\psi_1,\psi_2)$. 
The Majorana spinor satisfies
\begin{align}
\bar{\psi}=\psi^{{T}}C=\psi^\dagger D,\quad C=D=i\sigma_2
\end{align} where the charge conjugation matrix defined by $C\Gamma^\mu=- (\Gamma^\mu)^T C$ and the Dirac conjugation matrix $D\Gamma^\mu=- (\Gamma^\mu)^\dagger D$ are both given by the Pauli matrix $\sigma_2$. 
Then the Majorana condition is satisfied by the reality condition $\psi^{*}=\psi$, and the action of the free Majorana fermion model with Carrollian symmetry\footnote{There will be an overall minus sign in front of the action for the Galilean case.} can be written as~\cite{Hao:2022xhq,Yu:2022bcp}
\begin{align}
    S&=-\frac{i}{8\pi}\int dx^{\rho}\land dx^{\nu}\epsilon_{\rho\nu}\bar{\psi}\Gamma^{\mu}\partial_{\mu}\psi\label{fermionaction}\\
  & = -\frac{i}{4\pi}\int dxdy(2\psi_1 \partial_x\psi_1-\psi_2\partial_y\psi_1-\psi_1\partial_y\psi_2).\nonumber
\end{align}
Using the standard Neother procedure, the stress tensor is given by \cite{Hao:2022xhq,Yu:2022bcp}
\begin{equation}
T=-\frac{i}{2}(\psi_1\partial_x\psi_2+\psi_2\partial_x\psi_1),\quad M=-\frac{i}{2}\psi_1\partial_y\psi_2=-i\psi_1\partial_x\psi_1 ,\label{TM}
\end{equation}
where we have used the equation of motion in the second expression of $M$.
The spinor $\psi$ forms a primary multiplet with rank $2$ and conformal weight $1/2 $  in the highest weight representation. 
The $x$-ordered two-point correlation functions in highest weight vacuum read \cite{Hao:2022xhq,Yu:2022bcp}
\begin{align}\label{2pfh}
\langle \psi_1(x^\mu_{1})\psi_1(x^\mu_{2})\rangle &= 0,\nonumber\\
\langle\psi_1(x^\mu_{1})\psi_2(x^\mu_{2}) \rangle &= \frac{-i}{x_{12}},\quad x_{12}\equiv x_1-x_2\\
\langle\psi_2(x^\mu_{1})\psi_2(x^\mu_{2}) \rangle &= \frac{2i y_{12}}{x_{12}^{2}},\quad y_{12}\equiv y_1-y_2\nonumber
\end{align}
where the subscript of $x^\mu_i$ denotes the position of the operator, and the superscript denotes the two components of the coordinates. In the induced vacuum, time-ordering corresponds to ordering in $y$, and the two-point correlation functions read \cite{Hao:2022xhq}
\begin{align}\label{2pfI}
   \langle \psi_1(x^\mu_{1})\psi_1(x^\mu_{2})\rangle &= 0,\nonumber\\
\langle \psi_1(x^\mu_{1})\psi_2(x^\mu_{2})\rangle &= 2\pi \delta(x_{1}-x_{2})[\theta(y_{1}-y_{2})-\theta(y_{2}-y_{1})],\\
\langle \psi_2(x^\mu_{1})\psi_2(x^\mu_{2})\rangle &= 4\pi \delta'(x_{1}-x_{2})[y_{1}\theta(y_{1}-y_{2})+y_{2}\theta(y_{2}-y_{1})],\nonumber
\end{align}
The central charges are $c_{L}=1,c_{M}=0$ in the highest weight vacuum \cite{Hao:2022xhq,Yu:2022bcp} and $c_{L}=c_{M}=0$ for a particular induced vacuum \cite{Hao:2022xhq}.

\section{Uniformization map and modular Hamiltonian}\label{section3}
In this section, we use the uniformization map to find the modular Hamiltonian in the BMS free fermion theory. We first consider the single interval case to set up our uniformization map between various coordinate systems. As a consistency check, we show that the modular Hamiltonian from the uniformization map is the same as that obtained from the generalized Rindler transformation in \cite{Castro:2015csg,Jiang:2017ecm,Apolo:2020qjm}. 
Then the uniformization map can be generalized to the double-interval case.
For the free fermion model, by further introducing auxiliary fields, we compute the modular Hamiltonian for a region with two disjoint intervals. The result contains a local piece as well as a non-local piece.

Our strategy of finding the modular Hamiltonian for the BMS free fermion theory on an arbitrary subregion $\mathcal A\subset \mathcal M$ consists of three steps. The first step is to find a uniformization map which maps the interval $\mathcal A$ to a half interval, on which the modular Hamiltonian is geometrical and can be written as a conserved charge which generates rotational symmetry about the origin. The second step is to determine the modular Hamiltonian on the original plane by the transformation law of BMS symmetry. This enables us to express the modular Hamiltonian in terms of the coordinates and the stress tensor multiplet. For explicit models, the final step is to express the stress tensor in terms of fundamental fields.
The stress tensor contains a local term that comes from the transformation of a half-line interval under the uniformization map.
In general, the uniformization map is not a one-to-one map. If the interval contains $n$ disjoint segments, the inverse map contains  $2(n-1)$  branching points, around which twisted boundary conditions should be imposed on the fields. 
Similar to the free fermion CFT \cite{Wong:2018svs},  the twisted boundary conditions in G/C CFT can be implemented by the background field method that adds an additional term to the stress tensor, which will introduce a nonlocal term in the modular Hamiltonian. 
Putting pieces together, we will obtain an explicit expression for the modular Hamiltonian of two disjoint intervals.

\subsection{Uniformazition map}

Consider the G/C-CFT on a state specified by the density matrix $\rho$. Then the density matrix on an interval $\mathcal A$ is given by taking  partial trace over states on the complement $\bar{\mathcal A}$ \begin{align}
    \rho_{\mathcal A}=\tr_{\bar{\mathcal A}} \rho=e^{- \mathcal H},
\end{align} where $\mathcal H$ is the modular Hamiltonian.
The modular Hamiltonian generates a symmetry transformation, under which an operator flows as
\begin{equation}
\mathcal{O}(s,x^\mu)=e^{-i\mathcal H s}\mathcal{O}(x^\mu)\,e^{i\mathcal{H}s}
\end{equation}
so that the flowed operators satisfy the following differential equations,
\begin{equation}\label{floweq-general}
\partial_s \mathcal{O}(s,x^\mu)=-i[\mathcal H, \mathcal{O}(s,x^\mu) ].
\end{equation}
A modular Hamiltonian satisfies the following KMS relation
\eq{
   \tr e^{-\mathcal H} \mathcal O(s,x^\mu) &= \tr e^{-\mathcal H} \mathcal O(x^\mu)\\
\tr e^{-\mathcal H} \mathcal O_1(i,x_1^\mu)\, O_2(x_2^\mu)  &= \tr e^{-\mathcal H} \mathcal O_2(x_2^\mu)\, O_1(x_1^\mu). \label{eq:kms-2pt}
}
In general, the modular Hamiltonian is very complicated and most likely non-local. 
In some special cases, 
the modular Hamiltonian generates a geometrical flow and can be written as the associated Noether charge. In such cases, 
 the modular Hamiltonian  can be written covariantly as \cite{Ciambelli:2018ojf,Barnich:2012rz}
\begin{equation}
    \mathcal H=-i\int_{\mathcal A}dx\sqrt{h}\,
     T^{y}_{\ \ \mu}\xi^\mu 
\end{equation}
where $\xi$ is geometrical  modular flow generator, $h=\det(h_{ab})=1$ is the determinant of the metric $h_{ab}$, $a,b=x$ on $\mathcal{A}$, and  $T^{\mu}_{\ \ \nu}$ is the stress tensor. 

For BMSFT, the modular Hamiltonian for the highest weight vacuum state on the half-line interval ${ \mathcal {\hat R}}^+=\{(\hat z, \hat w)| \hat z\in[0,\,\infty),\hat w=0 \}$ with no displacement in the $\hat w$ direction is the same as that of a chiral CFT in two dimensions \cite{Bisognano:1976za}.  In this case, the modular Hamiltonian
generates boost symmetry in the Lorentzian signature, or rotational symmetry in the Euclidean theory with generator $\xi=-i{\hat z}\partial_{\hat z}-i\hat w \partial_{\hat w}$.  
As a result, the modular Hamiltonian is given by 
\begin{equation}
\mathcal H=-\int_{0}^{\infty} {\hat z} \,{\hat T}({\hat z},\hat w=0)\,d{\hat z}=-\int_{0}^{\infty} {\hat z} \,{\hat L}({\hat z})\,d{\hat z}.\label{Hzhat}
\end{equation}
This is the starting point for our subsequent discussions. 
Before moving on, let us comment on the modular boost generator from the perspective of taking the UR limit.
The causal development of a segment in BMSFT is a vertical strip, which is left invariant by $\xi$. In a CFT$_2$, the modular boost is given by $\xi^{\text{CFT}}=-iz\partial_z+i\bar z\partial_{\bar z}=i(l_0^+- l_0^-)$,
where $l_0^\pm$ is the left/right-moving Virasoro generator. Under the UR limit on the plane \cite{Hao:2021urq}, we have $l_n^+ -l_{-n}^-=l_n$ which becomes the Virasoro generators in the BMS$_3$ algebra. Therefore, the UR limit of the modular boost of a CFT$_2$ is precisely the BMS boost generator $\xi$, which is consistent with the intrinsic analysis from the Rindler method \cite{Apolo:2020bld}. 
Starting from the modular Hamiltonian on the half-line interval \eqref{Hzhat}, we can use BMS transformations to study other intervals.
If an interval $\mathcal A$ can be mapped to the half-line interval ${ \mathcal {\hat R}}^+$ by a BMS transformation, 
\begin{equation}
    \hat z=f(x),\quad \hat w=f'(x)\,y+g(x),\label{tilde2xy}
\end{equation}
then with \eqref{Tuv}, the modular Hamiltonian can be written schematically as
\eq{\mathcal H&=
-\int_{{V} } dx\left(X L(x)+Y M(x)\right)+const.
\label{Hg}
} 
where $V$ is the range of $x$ in the interval ${\mathcal A} $, $L$ and $M$ are components of the stress tensor, and the functions $X$ and $Y$ are determined by the transformation rule \eqref{transf},
\begin{equation}
X=f'^{-1}f,\quad Y=\frac{gf'-fg'}{f'^{2}}.\label{XY}
\end{equation}
The constant term in \eqref{Hg} is related to the anomalies and will affect the normalization of the density matrix. Nevertheless, the constant part is irrelevant in our subsequent discussions and will be ignored henceforth.
The modular Hamiltonian is then determined by the functions $X,Y$, and the stress tensor $L,M$. 

Let us comment more on the functions $X,Y$. Given a modular Hamiltonian on an interval $\mathcal A$ in the form of \eqref{Hg} with a specific choice of $X$ and $Y$, a further BMS transformation keeps the structure of \eqref{Hg} but modifies the functions $X$ and~$Y$. \linebreak[100]
A general BMS transformation \eqref{transf} consists of an $f$ transformation, which is sometimes referred to as superrotations, and the supertranslations parameterized by the function $g$. Let us now consider them separately.  
The $g$ transformation will change the $y$ coordinates of the end points but does not affect the $x$ coordinates, and will only affect the $Y$ function, 
\begin{equation}
y\to y+g(x),\qquad X\rightarrow X,\quad Y\rightarrow Y-2g'X+\partial_x(gX).
\label{gtrans}
\end{equation}
Consider the transformations $f$ and $g$ separately. 
The $f$ transformation reparametrizes the $z$ coordinate and is similar to the uniformization map in chiral CFT. 
Under an $f$ transformation, we have
\begin{equation}
x\to f(x),\qquad X\rightarrow f'^{-1}X,\quad Y\rightarrow f'^{-1}Y.\label{ftrans}
\end{equation}
If the transformation \eqref{tilde2xy} is single valued, the stress tensor in the $x$-$y$ space is completely determined from that in the $\hat z$-$\hat w$ space by the transformation rule \eqref{transf}. However, if the BMS transformation \eqref{tilde2xy} is multivalued, it will introduce twisted boundary conditions to the fields, or equivalently, new contributions to the stress tensor. We will return to this point momentarily.  
In the following, we will first derive the modular Hamiltonian on a single interval, which can be related to the half-line interval by a single-valued uniformization map.

\subsubsection*{Tilted half-line intervals }
On the half line ${\mathcal{\hat R}}^+$, we have
\begin{equation}
X=\hat z, \ \ Y=0.
\end{equation}
A generic half-line interval $\mathcal R^+$ can be generated by
a global $g$ transformation, 
  \begin{equation}
 z=\hat z,\quad w=\hat w+k_{+}\hat z^2+k_0\hat z+k_-,\label{kc}
  \end{equation}
which does not change the range of $z$, but changes the $w$ coordinates of the end points. 
Using the general rules \eqref{gtrans},  the modular Hamiltonian for $\mathcal R$ is
\begin{equation}\label{xys}
X=z,\quad Y= k_+ z^2-k_-.
\end{equation}
We find that the coefficients $k_0$ do not affect the result of the modular Hamiltonian.  For simplicity, we take $k_0=0$ henceforth.

As the transformation is globally well defined, it will not introduce additional singularities or branch cuts, so that stress tensor components $L$ and $M$ can be obtained from those on $\mathcal {\hat R}^+$  by the transformation rules \eqref{transf}. Therefore, the computation of the modular Hamiltonian on $\mathcal { R}^+$  parameterized by $k_\pm$ is complete.

In the case of free G/C fermion theory, the stress tensor is given by \eqref{TM}, so that the modular Hamiltonian is given by
\begin{align}
\mathcal H=&{i\over2}\int_{V} dz \Big( z \psi_1\partial_z(\psi_2-2w\partial_z\psi_1)+z(\psi_2-2w\partial_z\psi_1)\partial_z\psi_1+2(k_+z^2-k_-) \psi_1 \partial_z \psi_1\Big) \label{Hsinglezw}
\end{align}
where $V=[0,\infty)$. Despite the explicit dependence of the $w$ coordinate in the above expression, the modular Hamiltonian is indeed conserved after using the equation of motion.

Now, let us consider the modular flow.
Using the canonical anti-commutation relation
\begin{equation}\label{anticm}
\{\psi_1(z_1,w),\psi_2(z_2,w)\}=2\pi \delta(z_1-z_2),\ \ \{\psi_1,\psi_1\}=\{\psi_2,\psi_2\}=0,
\end{equation} 
the modular flow equation \eqref{floweq-general} becomes 
\begin{equation}
\partial_s \psi_1(s,z)=2\pi z\partial_z \psi_1 (s,z).
\end{equation}
In the $(z,\,w)$ plane,  the above equation has the following characteristic equation 
\begin{equation}\label{chareq}
\partial_s z=2\pi z
\end{equation}
which is exactly the geometric flow equation with the solution of the characteristic trajectory
\begin{align}
\label{chartraz}
    \psi_1(s,z)&=\psi_1(z(s)), \quad z(s)=z_0 e^{2\pi s}.
\end{align}
Note that the modular image of the fundamental field $\psi_1$ is $y$-independent, so that one cannot see the flow in the $y$ direction of $\psi_1$ field. 
Similarly, the modular flow equation of $\psi_2$ 
\begin{equation}
    \partial_s\psi_2=2\pi z\partial_z\psi_2+2\pi w\partial_w\psi_2+2\pi(k_+z^2-k_-)\partial_w\psi_2
\end{equation}
has the characteristic equation
\begin{equation}
    \partial_s z=2\pi z,\ \ \partial_s w=2\pi w+2\pi(k_+z^2-k_{-}).
\end{equation}
The solution of which reads,
\begin{equation}\label{chartra}
    w(s)=k_-+k_+ z(s)^2+c_1
   e^{2 \pi  s}
\end{equation}
where $c_1$ is a constant determined by the initial conditions.

As a result, the modular flow is completely determined by a geometrical flow
\begin{align}
    \psi_a(s;z,w)=\psi_a(z(s),w(s))
\end{align}
where the flowed coordinate $(z(s),w(s))$ is given by \eqref{chartraz} \eqref{chartra}.
\subsubsection*{Single intervals with finite length }
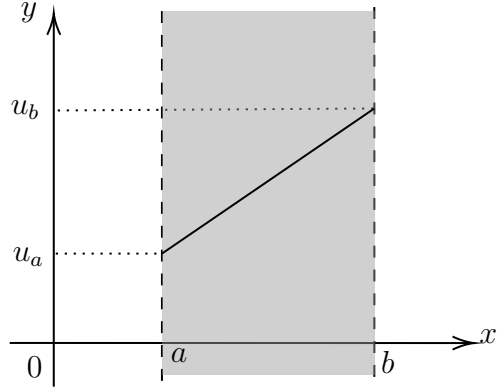
\begin{figure}[htbp]\centering
\begin{tikzpicture}[x=0.75pt,y=0.75pt,yscale=-1,xscale=1]
\draw    (179,192) -- (409,192.06) ;
\draw [shift={(411,192.06)}, rotate = 180.02] [color={rgb, 255:red, 0; green, 0; blue, 0 }  ][line width=0.75]    (10.93,-3.29) .. controls (6.95,-1.4) and (3.31,-0.3) .. (0,0) .. controls (3.31,0.3) and (6.95,1.4) .. (10.93,3.29)   ;
\draw    (201,214) -- (201,28.06) ;
\draw [shift={(201,26.06)}, rotate = 90] [color={rgb, 255:red, 0; green, 0; blue, 0 }  ][line width=0.75]    (10.93,-3.29) .. controls (6.95,-1.4) and (3.31,-0.3) .. (0,0) .. controls (3.31,0.3) and (6.95,1.4) .. (10.93,3.29)   ;
\draw  [dash pattern={on 4.5pt off 4.5pt}]  (255,25.06) -- (255,212) ;
\draw  [dash pattern={on 4.5pt off 4.5pt}]  (361,23.06) -- (361,210) ;
\draw  [dash pattern={on 0.84pt off 2.51pt}]  (202,147.06) -- (255,147.06) ;
\draw  [dash pattern={on 0.84pt off 2.51pt}]  (203,75.06) -- (361,74.06) ;
\draw  [draw opacity=0][fill={rgb, 255:red, 155; green, 155; blue, 155 }  ,fill opacity=0.47 ] (255,25.06) -- (361,25.06) -- (361,208.06) -- (255,208.06) -- cycle ;
\draw    (361,74.06) -- (255,147.06) ;

\draw (186,197.4) node [anchor=north west][inner sep=0.75pt]    {$0$};
\draw (258,193.4) node [anchor=north west][inner sep=0.75pt]    {$a$};
\draw (363,194.4) node [anchor=north west][inner sep=0.75pt]    {$b$};
\draw (179,142.4) node [anchor=north west][inner sep=0.75pt]    {$u_{a}$};
\draw (178,68.4) node [anchor=north west][inner sep=0.75pt]    {$u_{b}$};
\draw (412,183.4) node [anchor=north west][inner sep=0.75pt]    {$x$};
\draw (183,19.4) node [anchor=north west][inner sep=0.75pt]    {$y$};

\end{tikzpicture}
\caption{The configuration of a finite single interval. The shaded strip indicates the causal domain of dependence of the interval.}
\label{fig1}
\end{figure}

Now consider finite intervals, as shown in figure \ref{fig1}, with endpoints $(a,\,u_a)$ and $(b,\,u_b)$, which can be obtained from $\hat {\mathcal R}^+$ by a two-step BMS transformation: The first step is to tilt the half-line interval by a global BMS transformation in the form of \eqref{kc}, and the second step is a uniformazition map generated by a $f$ transformation,
\begin{equation}
z=\frac{x-a}{b-x},\quad w={(b-a)\over(x-b)^2}\,y.\label{z2x}
\end{equation}
Combining the two transformations, we learn that the relation between the parameters in \eqref{kc} and the end points of $\mathcal A$ is given by 
\begin{equation}
 k_+={u_b\over l},\quad k_0=-{u_a+u_b\over l},\quad k_-={u_a\over l},\quad 
l\equiv b-a. 
\end{equation}
Then the modular Hamiltonian is given by \eqref{Hg} with  
\begin{align}
X&=\big({dz\over dx}\big)^{-1}z=\frac{(x-a)(x-b)}{a-b},\label{XYfunction}\\
Y&=\big({dz\over dx}\big)^{-1}(k_+z^2-k_-)=-\frac{u_a(x-b)^2-u_b(x-a)^2}{(a-b)^2}.\nonumber
\end{align}
The above expression reproduces the result \eqref{RindlerH} which was obtained using the generalized Rindler method \cite{Apolo:2020qjm}.
In the free fermion theory, we have 
\begin{align}
\mathcal H=&{i\over2}\int_{V} dx \big({dz\over dx}\big)^{-1}\Big( z [\psi_1\partial_x(\psi_2-2y\partial_x\psi_1)+(\psi_2-2y\partial_x\psi_1)\partial_x\psi_1] +2(k_+z^2-k_-) \psi_1 \partial_x \psi_1\Big) \label{Hsingle}
\end{align}
where the uniformization map is given in \eqref{z2x}. We observe that the modular Hamiltonian is independent of $y$ after using the equation of motion. Note that the result for the tilted half-line interval \eqref{Hsinglezw} can also be written in the form of \eqref{Hsingle}, with $z=x,\,y=w$.
In fact, \eqref{Hsingle} is a general expression of the local term of the modular Hamiltonian, which generates a geometric flow via \eqref{floweq-general}. 
The flow trajectory can be directly computed in the $(x,y)$ plane, or equivalently by transforming the flow trajectory \eqref{chartraz} and \eqref{chartra} using the map \eqref{z2x}. The resulting trajectory is 
\begin{align}\label{flowxy1}
x(s)&=b-\frac{(a-b)(b-x_{0})}{e^{2\pi s}(a-x_{0})+x_{0}-b}.\\
y(s)&=k_+(x_{0}-b)^{-1} (a (x-x_0)+x
   (x+x_0))\nonumber\\
   &+k_- (a-x_0)^{-1} (b (x-x_0)+x
   (x+x_0))\nonumber\\
   &+(a-x_0)^{-1}(x_{0}-b)^{-1}y_0 (a (x-b)+x (x+b))
\end{align}
where $x=x(s)$ in the expression of $y(s)$.

\begin{figure}[htbp]\centering
\includegraphics[width=0.5\linewidth]{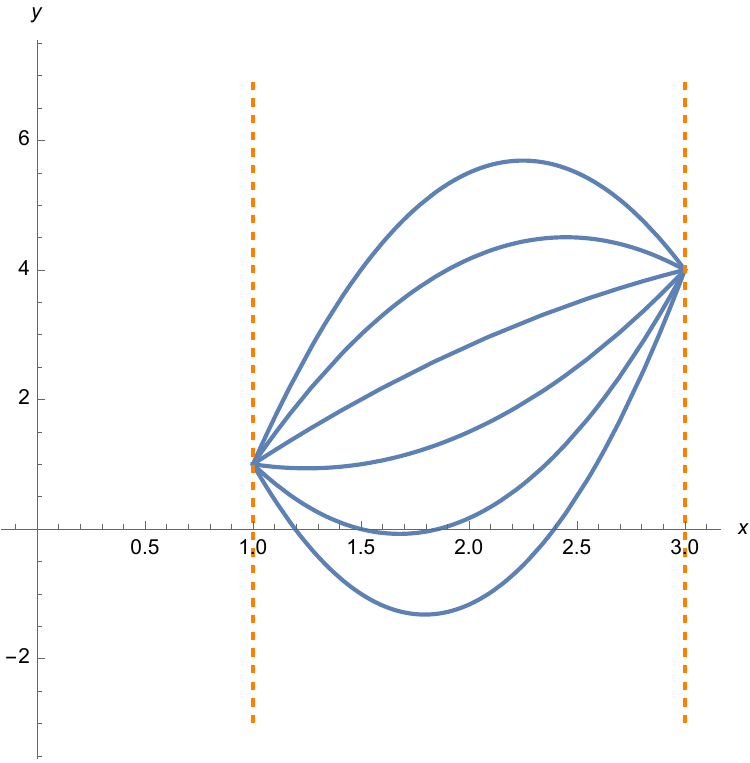}
\caption{The modular flow trajectory of a single interval, where we have chosen the parameters $a=1,b=3,k_{+}=2,k_{-}=0.5$.}
\label{fig2}
\end{figure}

To conclude this subsection, the modular flow in the single interval case is purely geometric, as shown in figure \ref{fig2}. The modular hamiltonian is local and can be written in the form of \eqref{Hsingle}, where the information of the interval is encoded in the BMS transformation between $(z,w)$ and $(x,y)$. 
This is, however, no longer the full story for the multi-interval cases,  as we will show explicitly in the following. 

\subsection{Two disjoint intervals}\label{section2}

\begin{figure}[!ht]\centering
\begin{tikzpicture}[x=0.75pt,y=0.75pt,yscale=-1,xscale=1]

\draw    (179,192) -- (590,194.05) ;
\draw [shift={(592,194.06)}, rotate = 180.29] [color={rgb, 255:red, 0; green, 0; blue, 0 }  ][line width=0.75]    (10.93,-3.29) .. controls (6.95,-1.4) and (3.31,-0.3) .. (0,0) .. controls (3.31,0.3) and (6.95,1.4) .. (10.93,3.29)   ;
\draw    (398,212) -- (398,84.06) -- (398,26.06) ;
\draw [shift={(398,24.06)}, rotate = 90] [color={rgb, 255:red, 0; green, 0; blue, 0 }  ][line width=0.75]    (10.93,-3.29) .. controls (6.95,-1.4) and (3.31,-0.3) .. (0,0) .. controls (3.31,0.3) and (6.95,1.4) .. (10.93,3.29)   ;
\draw  [dash pattern={on 4.5pt off 4.5pt}]  (255,25.06) -- (255,212) ;
\draw  [dash pattern={on 4.5pt off 4.5pt}]  (361,23.06) -- (361,210) ;
\draw  [dash pattern={on 0.84pt off 2.51pt}]  (398,148.03) -- (255,147.06) ;
\draw  [dash pattern={on 0.84pt off 2.51pt}]  (398,74.06) -- (361,74.06) ;
\draw  [draw opacity=0][fill={rgb, 255:red, 155; green, 155; blue, 155 }  ,fill opacity=0.47 ] (255,25.06) -- (361,25.06) -- (361,208.06) -- (255,208.06) -- cycle ;
\draw    (361,74.06) -- (255,147.06) ;
\draw  [dash pattern={on 4.5pt off 4.5pt}]  (435,18.13) -- (435,205.06) ;
\draw  [dash pattern={on 4.5pt off 4.5pt}]  (541,20.06) -- (541,207) ;
\draw  [dash pattern={on 0.84pt off 2.51pt}]  (399,121.06) -- (435,121.06) ;
\draw  [dash pattern={on 0.84pt off 2.51pt}]  (399,91.06) -- (541,91.06) ;
\draw  [draw opacity=0][fill={rgb, 255:red, 155; green, 155; blue, 155 }  ,fill opacity=0.47 ] (435,22.06) -- (541,22.06) -- (541,205.06) -- (435,205.06) -- cycle ;
\draw    (541,91.06) -- (435,121.06) ;

\draw (382,194.4) node [anchor=north west][inner sep=0.75pt]    {$0$};
\draw (258,193.4) node [anchor=north west][inner sep=0.75pt]    {$-b$};
\draw (340,192.4) node [anchor=north west][inner sep=0.75pt]    {$-a$};
\draw (373,127.4) node [anchor=north west][inner sep=0.75pt]    {$y_{-a}$};
\draw (374,53.4) node [anchor=north west][inner sep=0.75pt]    {$y_{-b}$};
\draw (598,187.4) node [anchor=north west][inner sep=0.75pt]    {$x$};
\draw (383,16.4) node [anchor=north west][inner sep=0.75pt]    {$y$};
\draw (438,193.4) node [anchor=north west][inner sep=0.75pt]    {$a$};
\draw (527,195.4) node [anchor=north west][inner sep=0.75pt]    {$b$};
\draw (377,79.4) node [anchor=north west][inner sep=0.75pt]    {$y_{b}$};
\draw (378,107.4) node [anchor=north west][inner sep=0.75pt]    {$y_{a}$};

\end{tikzpicture}
\caption{The configuration of two disjoint intervals.}
\label{fig3}
\end{figure}
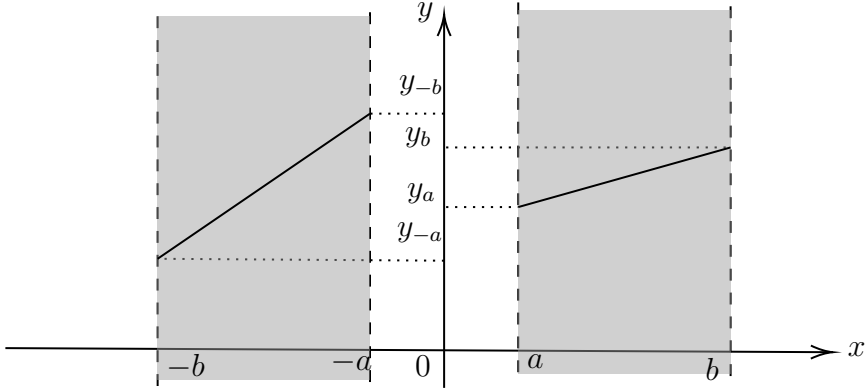

Two disjoint intervals can be obtained from the tilted half-line interval ${\mathcal { R}}^+$ specified by two parameters $k_\pm$ \eqref{kc}, as shown in figure \ref{fig3}.

An $f$ transformation in the form of,
\begin{equation}\label{unif2}
z=-\frac{(x+b)(x-a)}{(x+a)(x-b)},\ \ w=\frac{2 (b-a) \left(a
   b+x^2\right)}{(a+x)^2 (b-x)^2} y
  ,\quad b>a>0
\end{equation} maps 
 a two disjoint intervals region
 $\mathcal A={\mathcal A}_{1}\cup {\mathcal A}_{2}$ to ${\mathcal { R}}^+$. The $x$ coordinates of the end points have been chosen to be symmetric about the origin, determined by two parameters $a$ and $b$, so that the $x$ coordinates in the two segments satisfy \begin{align}\label{xends}
x_1\in V_1\equiv [-b,\,-a], \quad  x_2\in V_2\equiv  [a,b].
\end{align}
The $y$ coordinates at the four end points are determined by combining the transformations \eqref{unif2} and \eqref{kc},
\begin{align} \label{yends}
{y_{-b}\over bk_- }&={y_{-a}\over ak_+ }={y_a\over ak_- }={y_b\over bk_+ }=2{a-b\over a+b}.
\end{align}
From the above expression, we learn that the four $y$ coordinates of the end points are not arbitrary, but satisfy the following relations
\begin{equation}\label{yrelation}
{y_{-b}y_{-a}\over y_ay_b}=1,\quad {y_{-a}-y_a \over y_{-b}-y_{b} }=- {b\over a}.  
\end{equation}
In other words,  the two disjoint intervals we consider can be parameterized by $a,\,b,\,y_a,\,y_b$, while the $x$ coordinates are chosen to be symmetric about zero, and the $y$ coordinates satisfy the special relation \eqref{yrelation}.  
One can check that the cross ratios of the interval we have chosen are still arbitrary, 
\begin{equation}
{x_{34}x_{12}\over x_{13}x_{24}}=\frac{(a-b)^2}{(a+b)^2},\qquad {y_{12}\over x_{12}}+{y_{34}\over x_{34}}-{y_{13}\over x_{13}}-{y_{24}\over x_{24}}=\frac{8ab(k_--k_+)}{(a+b)^2},
\end{equation}
Therefore, more general cases of two-disjoint intervals can be related to the one specified by the end points \eqref{xends} \eqref{yends} by further global BMS transformations. In the following, we will focus only on \eqref{xends} \eqref{yends} without loss of generality. 

It is also useful to map the tilted half-line interval $\mathcal R^+$ back to the untilted one ${\hat {\mathcal R}}^+$ and express the parameters $k_\pm$ in terms of the end points, i.e.
\begin{align}
    k_+={y_b\over 2b}{a+b\over a-b},\quad k_-={y_a\over 2a}{a+b\over a-b}.\label{kc2}
\end{align}

The uniformization map \eqref{unif2} will determine $X$ and $Y$ in the modular Hamiltonian. 
However, as mentioned earlier,  the stress tensor in \eqref{Hg} will be modified if the map is not one-to-one. Unlike the case of a single interval, the inverse of the uniformization map of the double interval \eqref{unif2} is double-valued, with two branches:
\begin{align}\label{x12}
x_{1,2}&=-\frac{(a-b) (z-1) \pm(a+b)\sqrt{(z-z_+)(z-z_-)}}{2 (z+1)},\\ 
z_{\pm}&=\Big(\frac{\sqrt{b}\pm i\sqrt{a}}{\sqrt{b}\mp i\sqrt{a}}\Big)^2 \nonumber
\end{align}
For $z\in [0,\infty),$ we have 
 $x_{1}\in V_{1}\equiv [x_1, x_2]$ with the plus sign,  and $x_{2}\in V_{2} \equiv [x_3, x_4]$ with the minus sign in the above expression. 
If we extend the $f$ transformation to the entire complex plane, the transformation \eqref{unif2} introduces a new branch-cut in the complex $z$ plane, with branching points  $z_\pm$, at which the pre-image $x$ takes values of  
\begin{equation}
x_{1}=x_{2}=x_{\pm}=\pm i\sqrt{ab}.
\end{equation}

In the following, we will determine the expression of the stress tensor in the presence of the above branching points in the free G/C conformal fermion model.

\subsubsection{Bilocal modular Hamiltonian}

As explained in \cite{Casini:2009vk, Wong:2018svs}, the double-valued mapping \eqref{x12} can be made one-to-one if we make a two-fold cover of the complex $z$ plane $\Sigma_2$, with the first sheet mapping to the first branch $x_{1}$, and the second sheet to $x_{2}$. 
Then we can make an orbifold from
$\Sigma_2$ by identifying points on the two sheets with the same value of $z$. The replicated fields $O^{1},O^{2}$ originally living on the two sheets of $\Sigma_2$ now live in a single sheet but with non-trivial monodromy conditions around the branch points $z_{\pm}$. In the free G/C theory \eqref{fermionaction}, the replicated fermion fields $\psi^{(i)}$ satisfy the conditions  \begin{align}
&\psi^{(1)}(e^{2\pi i} z_+)=\psi^{(2)}(z_+),\quad \psi^{(2)}(e^{2\pi i} z_+)=-\psi^{(1)}(z_+),\\
&\psi^{(1)}(e^{2\pi i} z_-)=-\psi^{(2)}(z_-),\quad \psi^{(2)}(e^{2\pi i} z_-)=\psi^{(1)}(z_-).
\end{align}
The above monodromy conditions can be diagonalized by
\begin{align}
\Psi_+=\frac{1}{\sqrt{2}}(\psi^{(1)}+ i\psi^{(2)})\equiv (\beta,\alpha)^T,\quad \Psi_-=(\bar \beta,\bar \alpha)^T.
\end{align}
For later convenience, we will refer to fermions $\psi^{(i)}\equiv(\beta_i,\,\alpha_i)^T$ as the replica basis and $\Psi_\pm$ as the diagonal basis. 
In the diagonal basis, the orbifold theory can be described by the free action 
\begin{align}\label{orbifoldaction}
S_{tw.b}&=-\frac{i}{4\pi}\int dzdw (\bar \Psi_+\Gamma^\mu\partial_\mu \Psi_++\bar \Psi_-\Gamma^\mu\partial_\mu \Psi_-)\\
&=-\frac{i}{4\pi}\int dzdw(2\bar{\beta}\partial_z\beta +2\beta\partial_z\bar{\beta} -\bar{\alpha}\partial_w\beta-\alpha\partial_w\bar{\beta}-\bar{\beta}\partial_w\alpha-\beta\partial_w\bar{\alpha})
\end{align}
with twisted boundary conditions
\begin{equation}\label{twistbc}
\Psi_\pm (e^{2\pi i} z_+)=\mp i\Psi_\pm ( z_+),\quad \Psi_\pm (e^{2\pi i} z_-)=\pm i\Psi_\pm ( z_-). 
\end{equation}
The action \eqref{orbifoldaction} describes two free G/C fermions, and has a global $U(1)$ symmetry which shifts the phase of $\Psi$ by a coordinate independent constant
\begin{equation}
    \Psi_\pm\to e^{\pm iq}\Psi_\pm.
\end{equation}
The Noether current of this $U(1)$ symmetry is given by 
\begin{align}
j^\mu&={1\over2}\Big(\bar \Psi_+\Gamma^\mu\Psi_+-\bar \Psi_-\Gamma^\mu\Psi_-\Big)\nonumber \\
j^z&=J_+-J_-, \quad J_+\equiv\bar{\beta}\beta=i\beta_1\beta_2, \quad J_-\equiv {\beta}\bar \beta=i\beta_2\beta_1\\
j^w&={1\over2}\Big(\beta \bar \alpha+\alpha \bar \beta-\bar{\beta}\alpha-\bar\alpha \beta\Big)=-{i\over2}\epsilon^{ij}\Big( \beta_i\alpha_j+ \alpha_i\beta_j\Big)
\nonumber 
\end{align}
where we have expressed the currents in both the diagonal basis and the replica basis. 
The global $U(1)$ transformation can be gauged by introducing a gauge field so that 
\begin{equation}
    \Psi_\pm \to e^{\pm i\theta(x)}\Psi_\pm,  
    \quad A_\mu\to A_\mu-\partial_\mu \theta ,\label{gauge0}
\end{equation}
Then the twisted boundary condition \eqref{twistbc} can be realized by a background gauge field
\begin{equation}
    \Psi_\pm \to e^{\pm i\int  A_\mu dx^\mu }\Psi_\pm,  
    \label{gauge}
\end{equation}
whose holonomy along contours $C_\pm$ around $z_\pm$ is given by 
\begin{align}
     e^{i\int_{C_\pm} A_\mu dx^\mu} =\pm i
\end{align}
where the integration is counter-clockwise.
The above holonomy condition can be solved by the background field 
\begin{equation}
A_z=A(z)=\frac{1}{4i}\left(\frac{1}{z-z_+}-\frac{1}{z-z_-}\right),\quad A_w=0. \label{bg}
\end{equation}
The background field $A_\mu$ behaves as a rank-2 doublet. Under the BMS transformation
\begin{equation}
     A_w\rightarrow(f')^{-1}A_w,\ \ A_z\rightarrow(f')^{-1}\left(A_z-\frac{wf''+g'}{f'}A_w\right).
\end{equation}
Note that $A_z,A_w$ behave as the lower and upper components of the doublet due to the property that $A_\mu$ is contravariant. Also, since $A_w=0$, the doublet truncates to a short representation where only the lower component is non-vanishing and behaves as a singlet under the transformation.

In terms of the fermions \eqref{gauge} with standard boundary conditions, the free action \eqref{orbifoldaction} with twisted boundary conditions can then be written as
\begin{align}\label{sat}
    S_{A}&=-\frac{i}{4\pi}\int dzdw \Big(\bar \Psi_+\Gamma^\mu (\partial_\mu+iA_\mu) \Psi_++\bar \Psi_-\Gamma^\mu (\partial_\mu-iA_\mu) \Psi_-\Big)
\end{align}
and the equations of motion are the parallel transport condition $\Gamma^\mu (\partial\pm i A_\mu)\Psi_\pm=0$.
The stress tensor now contains the free part
\begin{align}
L&=-\frac{i}{2}[\bar\beta\partial_z(\alpha-2w\partial_z\beta)+(\bar\alpha-2w\partial_z\bar\beta)\partial_z\beta+\beta\partial_z(\bar\alpha-2w\partial_z\bar\beta)+(\alpha-2w\partial_z\beta)\partial_z\bar\beta]\nonumber\\
&=-{i\over2} \sum_i[\beta_i\partial_z (\alpha_i-2w\partial_{z}\beta_{i})+(\alpha_i-2w\partial_{z}\beta_i)\partial_z\beta_i],\\ 
M&=-\frac{i}{2}(\bar\beta\partial_w\alpha+\beta\partial_w\bar\alpha)=-{i\over2} \sum_i\beta_i\partial_w \alpha_i\end{align}
and an additional part due to the presence of the background gauge field,
\begin{align}
L_i
&=-A (j^w+w\partial_z j^{z})\nonumber\\
&={A\over2}[\bar \beta (\alpha-2w\partial_z\beta) +(\bar\alpha-2w\partial_z \bar\beta)\beta-\beta (\bar \alpha-2w\partial_z\bar\beta) -(\alpha-2w\partial_z\beta) \bar  \beta]\nonumber\\
&={i A\over2}\epsilon^{ij} [\beta_i(\alpha_j-2w\partial_z\beta_j)+(\alpha_i-2w\partial_z\beta_i)\beta_j]
,\\
M_i&=A j^z = A(\bar\beta \beta-\beta\bar\beta)=iA \epsilon^{ij}\beta_i\beta_j.
\end{align}

From the above expression, we learn that the additional terms actually indicate interactions between fermions in the two branches of the Riemann surface. As a consistency check, we find that the total stress tensors with the background U(1) field are conserved,
\begin{align}
L_{A}&=L+L_{i},\quad M=M+M_{i},\label{LMA}\\
\partial_w M_A&=0, \quad\partial_w  ( T_A-w\partial_z  M_A)=\partial_{w} {L_A}=0.
\end{align}
The orbifold theory with the action \eqref{orbifoldaction} and the twisted boundary condition \eqref{twistbc} is equivalent to the action \eqref{sat} with the non-trivial background field \eqref{bg} but with a trivial boundary condition. 
Now, it is straightforward to calculate the modular Hamiltonian.  In the $(z,w)$ coordinate, the modular Hamiltonian can be written in terms of the general expression \eqref{Hg} with \eqref{xys} and the modified stress tensor,
\begin{equation}\label{Hzw}
\mathcal H=-\int_{0}^{\infty} (z  L_A+(k_+z^2-k_-)  M_A)\,dz
\end{equation}
where $k_\pm$ are specified by \eqref{kc2}.
The modular Hamiltonian contains a local term that comes from the free theory,
\begin{equation}
\mathcal H_{local}=-\int_{0}^{\infty} (z L+(k_+z^2-k_-) M)\, dz,
\end{equation}
and a bi-local term which comes from the interaction between the two replicated fermions, 
\begin{equation}
\mathcal H_{non}=i\int dz A\epsilon^{ij}\Big( z \beta_i(\alpha_j-2w\partial_z\beta_j)+(k_+z^2-k_-)\beta_i\beta_j\Big)
\end{equation}
where $\epsilon^{12}=-\epsilon^{21}=1$. The above expression has been put in the replicated basis, with all the fields on the tilted plane \eqref{kc} parameterized by $(z,w)$.

The modular Hamiltonian for two-disjoint intervals can then be obtained by transforming \eqref{Hzw}  back to the $(x,\, y)$ plane. Due to the double-valuedness of the map \eqref{unif2}, extra care should be taken with the transformation of fields as well as the explicit dependence of the coordinates.
Explicit dependence of $(z,w)$ in \eqref{Hzw} can be directly rewritten in terms of $(x,\, y)$ using the uniformization map \eqref{unif2}. 
Note that the two branches in \eqref{x12} are related by 
\begin{equation}
x_1 x_2=-ab \label{x12ab}
\end{equation}
which allows us to view $x_{2}$ as a function of $x_{1}$, or vice versa. 
In terms of the coordinates in the first branch, we have  
\begin{align}
dz&=\frac{dz}{dx_{1}}dx_{1},\ \ z=z(x_{1})
\\A dz &=A {dz\over dx_{1}}dx_1 =\sqrt{dx_{2}\over dx_{1} }{1\over x_{1}-x_{2}} dx_1
\end{align}
where in the second line we have plugged in the explicit expression of the background gauge field \eqref{bg}.
As discussed in the previous subsection, the replicated field $\psi_{i}(z,w)$ should be mapped to $\psi(x_i,y_i)$ in the $(x,y)$ coordinate where $x_{i}$ denotes the preimage of $z$ in the $i$-th branch of the inverse map \eqref{x12}.
More explicitly, we have 
\begin{equation}
\beta_i(z)=\left({dz \over dx_{i}}\right)^{-{1\over 2}}\psi_1(x_{i}),\quad \alpha_i(z,w)=\left({dz \over dx_{i}}\right)^{-{1\over 2}} \psi_2(x_{i},y_i)-\frac{y_i {d^2z \over dx^2_{i}}}{\big({dz \over dx_{i}}\big)^{\frac{3}{2}}}\psi_1(x_{i}).\label{z2xy}
\end{equation}
Plugging these into the expression of the modular Hamiltonian, we obtain the bi-local term
\begin{align}
\mathcal H_{non}={}&i\sum_{i\neq j}\int_{V_i} dx_i \frac{z(x_j)\psi_1(x_i)(\psi_2(x_j,y_j)-2y_j\partial_{x_{j}}\psi_1(x_j))}{x_i-x_j}\left(\frac{dz}{dx_j}\right)^{-1}\nonumber\\
&+i\sum_{i\neq j}\int_{V_i} dx_i \frac{(k_+z(x_i)z(x_j)-k_-)\psi_1(x_i)\psi_1(x_j)}{x_i-x_j}\left(\frac{dz}{dx_j}\right)^{-1}.\label{nonlocH2}
\end{align}
The local term takes the same form as \eqref{Hsingle}, but with the integration region $V=V_1\cup V_2$ and the uniformization map \eqref{unif2}.  
Similar to the CFT case \cite{Casini:2009vk}, the local term and the bilocal term can be combined into a more compact expression, 
\begin{align}
    \mathcal H&=i\int_{V} \int_{V} dxdx' \Big[\frac{\delta(\log z(x)-\log z(x'))}{2(x-x')}(\psi_1(x)\psi_2(x',y')+\psi_2(x,y)\psi_1(x'))\nonumber\\
    &+(y'-y)\,\partial_{x'}\left(\frac{\delta(\log z(x)-\log z(x'))}{(x-x')}\right)\psi_1(x)\psi_1(x')\nonumber\\
    &+\frac{k_+\delta( \frac{1}{z(x)}- \frac{1}{z(x')})-k_-\delta( z(x)- z(x'))}{x-x'}\psi_1(x)\psi_1(x')\Big]\label{Hdouble1}
\end{align}
where $V=V_1\cup V_2$ is the entire range of $x$ of the interval. The double integral in the above expression can be split into integration over $\sum_i V_i\times V_i$ and $\sum_{i\neq j}V_i\times V_j$, the first of which corresponds to the local part and the second to the non-local part \eqref{nonlocH2}.
To rewrite the local term in the bilocal form \eqref{Hdouble1}, we have used integration by parts and replaced $\delta'(x)$ by $-{\delta(x)\over x}$ inside the integral as the test function vanishes at $x=0$ due to anticommutativity of the fermions. 
We find that the first line of \eqref{Hdouble1} is the same as the chiral fermion in relativistic theories \cite{Casini:2009vk}.

Although the modular Hamiltonian \eqref{Hdouble1} was derived for two disjoint intervals, it applies to the single interval as well if we take the region $V=[a,b]$ and the uniformization map \eqref{unif2}. In this case, the only contribution in \eqref{Hdouble1} comes from the local term, which, after integration, goes back to \eqref{Hsingle}.
This suggests that the result \eqref{Hdouble1} can be generalized to multi-intervals as well.  Similar to the discussion in CFT$_2$ as explicitly spelled out in \cite{Wong:2018svs}, the modular Hamiltonian can indeed be generalized to a class of multiple intervals which can be mapped to the tilted half-line interval specified by $(k_+,k_-)$ by the uniformization map
    \begin{align}\label{un}
    z(x)\equiv -\prod^n_{i=1}{(x-a_i)\over (x-b_i)},\quad  w={dz\over dx}y
\end{align}
In this case, the interval corresponds to $x\in V_i$, $V_i=[a_i,\,b_i]$, $i=1,\cdots, n$, and the end-point values of $y$ are determined by the map \eqref{un} together with the parameters $k_+,k_-$.
We will come back to this point in the next section.

\subsubsection{Modular flow}
\begin{figure}[htbp]
    \centering
    \includegraphics[width=0.5\linewidth]{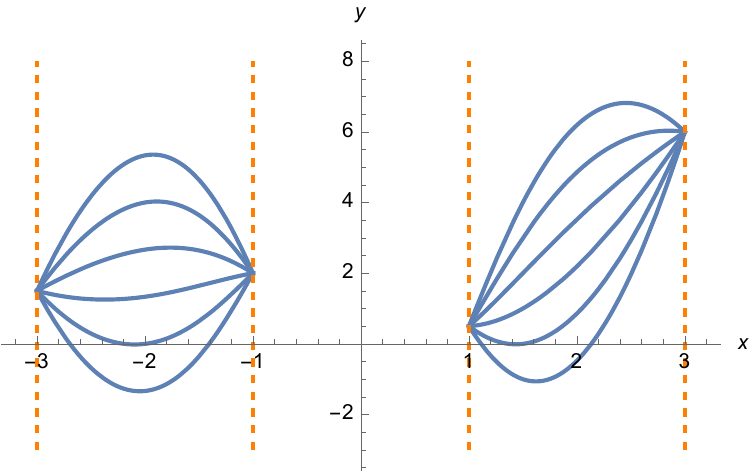}
    \caption{The geometric modular flow of two disjoint intervals, where we have chosen the parameters $a=1,b=3,k_{+}=2,k_{-}=0.5$.}
    \label{fig4}
\end{figure}

Now let us discuss the modular flow in the $z$ coordinate generated by the modular Hamiltonian \eqref{Hzw}. 
Using the commutation relations, we get 
\begin{align}\label{floweq}
\frac{d\beta(s,z)}{ds}&=2\pi z\, (\partial_z+iA )\,\beta(s,z),\\
    \frac{d\alpha(s,z,w)}{ds}&=2\pi z\,(\partial_z+iA )\,\alpha+2\pi w\,\partial_w\alpha+2\pi\left(k_{+}z^{2}-k_{-}\right)(\partial_z+iA )\,\beta.
\end{align}
Without the background field $A$, the solution of the above flow equation is a geometric flow $\beta(z(s))$, as shown in figure \ref{fig4}, where $z(s)$ is the solution of the characteristic equation  given by 
\begin{equation}\label{flowz}
    z=z_0e^{2\pi s}
\end{equation}

The effect of a non-vanishing background field is to introduce an additional phase factor, so that the solution of the flow equation \eqref{chartra} is given by
\begin{align}\label{flowd}
\beta(s,z)&=e^{i\theta_s}\beta(z(s)),\quad \\
\alpha(s,z,w)&=e^{i\theta_s}[\alpha(z(s),w(s))+ 
\lambda_s \beta(z(s))]
\end{align}
where
\begin{align}\label{thetas}
\theta_s&=\int_{z_0}^{z(s)} dz A={1\over4 i}\left(\log\frac{z(s)-z_+}{z(s)-z_-}-\log\frac{z_0-z_+}{z_0-z_-}\right)\\
\lambda_s&=i\int_{s=0}^s dz (k_+z-k_-z^{-1})A
\nonumber
\end{align}
Similarly, we have 
\begin{equation}
    \bar\beta(s,z)=e^{-i\theta_s}\bar \beta(z(s)).
\end{equation}
Compared to the single interval result \eqref{flowxy1} where the modular flow is purely geometrical, the modular flow for disjoint intervals has an additional phase shift. Similar to the result for relativistic fermions \cite{Casini:2009vk}, the BMS fermions also have an additional mixing between the two components of the spinor.
On the replica basis, this phase shift will introduce a mixing between the two replicas, so that we have 
\begin{align}\label{flowi}
    \beta_i(s,z)&=\Omega_{ij} {\beta}_j(z(s)),\quad \Omega=\left[
\begin{array}{cc}
 \cos \theta_s  & \sin \theta_s  \\
 \, -\sin \theta_s\, & \, \cos \theta_s \, \\
\end{array}
\right],\\
\alpha_i(s,z,w)&=\Omega_{ij}{\alpha}_j(z(s),w(s))+i\lambda_s \epsilon_{ij}\Omega_{jk} \beta_k(z(s)).
\end{align}
Finally, on the physical plane where the interval was originally defined, mixing between the two replicas becomes a mixing between the two branches of the uniformization map, i.e., points satisfying the relation \eqref{x12ab}.
To better describe the mixing,  
let us combine fields at $x_1$ and its image in the second branch $x_2=-ab/x_1$ into a column vector,  
whose images under the geometric flow are given by
\begin{align}\label{tilted}
    \tilde \beta (x_1(s))\equiv(\psi_1(x_1(s)), \, \psi_1(-ab/x_1(s)))^T
\end{align}
Using the uniformation \eqref{z2xy} and the solution \eqref{flowi}, the modular flowed fields in the $(x,y)$ plane are given by
\begin{align}
    \tilde \beta (s,x_1)&\equiv(\psi_1(s,x_1),\,\psi_1(s,-ab/x_1))^T={N }\Omega N^{-1}\,\tilde\beta(x_1(s))\\  \tilde{\alpha}(s,x_1,y)&\equiv(\psi_2(s,x_1,y),\,\psi_2(s,-ab/x_1,\frac{ab}{x_1^{2}}y))^T\\
    &={N }\Omega N^{-1}\,\tilde\alpha(x_1(s),y(s))+i\lambda_s {N }\epsilon\Omega N^{-1}\,\tilde\beta(x_1(s))\nonumber
\end{align}
where $\epsilon$ is the 2-dimensional anti-symmetric tensor, and $N$ is the transformation matrix due to the uniformation map
\begin{align}
    N_{ij}=\delta_{ij} \sqrt{dz\over dx_i}.\label{Nij}
\end{align}
Now the rotation matrix $N\Omega N^{-1}$ mixes fields in sub-regions $x_1\in V_1$ and $x_2\in V_2$, and thus introduces non-locality.  

As evident from our derivation, the non-locality originates from the additional phase shift in the diagonal basis \eqref{flowd}, which in turn can be understood as follows. 
The 1-form field $A_\mu$ transforms contravariantly under the BMS transformation, 
\begin{equation}\label{Atrs}
     A_w\rightarrow(f')^{-1}A_w,\quad A_z\rightarrow(f')^{-1}\left(A_z-\frac{wf''+g'}{f'}A_w\right)
\end{equation}
so that $dx^\mu A_\mu$ is BMS invariant. In other words, $A_z,\,A_w$ behave as lower and upper components of a BMS doublet. 
Recall that the fermions transform as 
\begin{equation}\label{psitrs}
\beta\rightarrow (f')^{-1/2}\beta,\quad \alpha\rightarrow (f')^{-1/2}\alpha-(f')^{-3/2}(wf''+g')\,\beta.
\end{equation}
It is not difficult to verify that the action  \eqref{sa} is invariant under the BMS transformations with the above transformation rules \eqref{Atrs} and \eqref{psitrs}.
However, what we really have in the action \eqref{sa} is a background field specified by \eqref{bg}
with vanishing $A_w$. 
First note that BMS transformations leave the sector with $A_w=0$ invariant, while $A=A_z$ transforms as 
\begin{equation}
A\rightarrow (f')^{-1} A
\end{equation}
which is similar to the transformation of a weight $1$, charge $0$ singlet. 
To keep the expression of the background field, we need to supplement the BMS transformation $z=f(x)$ by two gauge transformations \eqref{gauge0}, which undo the dressing in \eqref{gauge} before the BMS transformation, and re-introduce one afterwards. In the diagonal basis, we can regard the modular flow \eqref{flowz} as a BMS-$f$ transformation. Then the fermions acquire a total phase shift given by \eqref{flowd}.

\subsection{Modular correlation function}
In this section, we calculate the modular flow of the two-point correlation function \begin{equation}
    G_{ab}(s)\equiv\langle\psi_a(s;x,y)\psi_b(s';x',y')\rangle\label{Gs}.
\end{equation}
When $s=0$, the propagator in the highest NS vacuum is given by \eqref{2pfh}. To get the flowed version, we proceed as follows. 
First, we compute the correlators in the diagonal basis on the 
$(z,w)$ plane, where modular flow acts as a simple geometric flow accompanied by a phase shift. The modular correlator \eqref{Gs} is then obtained by transforming the flowed correlator back to the original 
$(x,y)$ plane.
In order to do so, it is convenient to introduce the angular coordinate 
\begin{align}\label{thz}
    \theta={1\over4 i}\log\left({z-z_+\over z-z_-}\right)-\theta_0,\quad  \tan \theta_0=\sqrt{a\over b}, \quad z_\pm=e^{\pm i 4\theta_0}
\end{align}
so that the phase shift in \eqref{flowd}
is given by \begin{align}\theta_s=\theta|_{z=z(s)}-\theta|_{z=z_0}.\end{align}
Using the inverse transformation \eqref{x12}, we obtain \begin{equation}
    x_{j}=\sqrt{ab}\tan(\theta-(j-1)\pi/2),\quad j=1,2, \label{xjth}
\end{equation}
where the subscript $j$ labels the two branches of pre-images in the $(x,y)$ plane that maps to the same point $(z,w)$.
Note that the two branches are completely incorporated into a constant phase shift and therefore the differential relation of \eqref{xjth}  is independent of the choice of the branch, i.e.
\begin{align}
    {d\theta\over dx_i}={\sqrt{ab}\over(x_i^2+ab)}=\frac{1}{\sqrt{ab}}{\cos^2(\theta-(j-1)\pi/2)}.
\end{align}
This is another advantage of using the angular coordinate $\theta$, as the above derivative will show up frequently in the coordinate transformation.
In the replica basis, we thus have the unflowed two-point function,  
\begin{equation}
\langle \beta_i(\theta)\alpha_j(\theta')\rangle \equiv \big(\frac{d\theta}{dx_i}\big)^{-{\frac{1}{2}}}\big(\frac{d\theta}{dx'_j}\big)^{-{\frac{1}{2}}}
\langle \psi_{1}(x_i)\psi_{2}(x'_j,y'_j)\rangle
=\frac{-i}{\sin\big(\theta-\theta'+\frac{(i-j)\pi}{2}\big)}.
\end{equation}
We have the flowed two-point function 
\begin{equation}
\langle \beta_i(s,\theta)\alpha_j(s',\theta')\rangle =\Omega_{ik}(\theta_s)\Omega_{jl}(\theta'_s)\langle \beta_k(\theta(s))\alpha_l(\theta'(s'))\rangle
\end{equation}
where $\Omega$ is the rotational matrix defined in \eqref{flowi}.
Further transforming back to the $(x,y)$ plane, we obtain
\begin{align}
\langle \psi_1(s,x_i)\psi_2(s',x'_j,y'_j)\rangle&=(\frac{d\theta(s)}{dx_i})^{\frac{1}{2}}(\frac{d\theta'(s')}{dx'_j})^{\frac{1}{2}}\langle\beta_i(s,\theta)\alpha_j(s',\theta')\rangle \nonumber\\
&={N_{ii}(s)N_{jj}(s')\over N_{kk}(s)N_{ll}(s')}\Omega_{ik}(\theta_s)\Omega_{jl}(\theta'_{s'})
\langle \psi_{1}(x_k(s))\psi_{2}(x'_l(s'),y'_l(s'))\rangle 
\end{align}
where summations over $i,j,k,l$ are assumed, and the transfer matrix $N_{ii}$ is defined in \eqref{Nij}. 
Similarly, 
\begin{align}
    \langle \psi_2(s,x_i,y_i)\psi_2(s',x'_j,y'_j)\rangle
=&\frac{N_{ii}N_{jj}}{N_{kk}N_{ll}}\Bigg[\Omega_{ik}(\theta_{s})\Omega_{jl}(\theta'_{s'})\left\langle \psi_{2}(x_{k}(s),y_{k}(s))\psi_{2}(x'_{l}(s'),y'_{l}(s'))\right\rangle \nonumber\\
&+2i\lambda_s\epsilon_{ik}\Omega_{ks}(\theta_{s})\Omega_{jl}(\theta'_{s'})\left\langle \psi_{1}(x_{s}(s),y_{s}(s))\psi_{2}(x'_{l}(s'),y'_{l}(s'))\right\rangle \Bigg].
\end{align}
In terms of $\theta$, the results can be written as 
\begin{align}
    \left\langle \psi_{1}(s,x)\psi_{2}(s',x',y')\right\rangle &=-2i\frac{\cos\theta(s)\cos\theta'(s')}{\sqrt{ab}}\frac{\cos[2(\theta(s)-\theta'(s'))-(\theta(0)-\theta'(0))]}{\sin(2(\theta(s)-\theta'(s')))},
\end{align}
\begin{align}
   \langle &\psi_{2}(s,x,y)\psi_{2}(s',x',y')\rangle
= 4\lambda_s\frac{\cos\theta(s)\cos\theta'(s')\sin[2(\theta(s)-\theta'(s'))-(\theta(0)-\theta'(0))]}{\sqrt{ab}\sin[2(\theta(s)-\theta'(s'))]}\nonumber\\   
   &-2i \frac{\cos\theta(s) \cos\theta'(s')
   \csc ^2(\theta(s)-\theta'(s')) \cos [\theta(s)-\theta'(s')-(\theta(0)-\theta'(0))](y'-y) }{\sqrt{a b}}\nonumber\\
   &-2i\frac{\sin\theta(s)\cos\theta'(s')\sec^{2}(\theta(s)-\theta'(s'))\sin [\theta(s)-\theta'(s')-(\theta(0)-\theta'(0))](y'-y\cot^{2}\theta(s))}{\sqrt{ab}}.
\end{align}
As a consistency check, we have verified that the modular correlators indeed satisfy the KMS relation \eqref{eq:kms-2pt}.

\section{Entanglement entropy}
In this section, we discuss the general proposals and ansatz on the modular Hamiltonian and derive the relation between the entanglement entropy and the 2-point function using the coherent state method.

We will see that the multi-interval entanglement entropy for 2D free-fermion CCFT is the same as that for 2D chiral CFT in the literature. This conclusion is consistent with the highest weight vacuum BMS algebra structure of the 2D free fermion CCFT model, where $c_{L} = 1$ and $c_{M} = 0$.

\subsection{Modular Hamiltonian in general}
As can be seen explicitly from the previous section, the modular Hamiltonian for a large class of intervals that can be mapped to the $(z,w)$ plane by \eqref{un} is described by the expression \eqref{Hdouble1}. 
The modular Hamiltonian \eqref{Hdouble1}   has several features: it is i) real, ii) self-adjoint, 
and iii) there are no quadratic terms for $\psi_2$. 
This observation motivates us to assume that these properties hold more generally. 
In the following, we take the following ansatz for the modular Hamiltonian  
\begin{align}
    {\mathcal H}=\int dx dx'\psi_{a}(x,y) H^{(ab)}(x,y;x',y')\psi_{b}(x',y'), \quad a,b=1,2, \quad H^{(22)}=0. \quad \label{Hstructure}
\end{align}
Using the anti-commutativity of the Grassmannian variables, we learn that the above ansatz implies that the kernels are antisymmetric in positions 
\begin{align}
  H^{(ab)}(x,y;x',y')=-H^{(ba)}(x',y';x,y).  \label{anti}
\end{align}
They also satisfy the reality condition  
\begin{align}
 H^{(ab)*}(x',y';x,y)=H^{(ba)}(x,y;x',y')
 =-H^{(ab)}(x',y';x,y)\label{real}
\end{align}
where we have used \eqref{anti} in the last equation. 
In addition, we also assume that each kernel is individually self-adjoint 
\begin{align}
    \quad H^{(ab)*}(x',y';x,y) =H^{(ab)}(x,y;x',y').\label{sa}
\end{align}
From \eqref{anti} and \eqref{sa} we also learn that
\begin{align}
    H^{(12)}(x,y;x',y')=H^{(21)}(x,y;x',y').
\end{align}
It is straightforward to check that the explicit result \eqref{Hdouble1} indeed satisfies the above conditions \eqref{anti}-\eqref{sa}.

To simplify the calculation, it is also useful to introduce the $y$ independent fields 
\begin{align}\label{bega}
   \hat\psi_1(x)=\psi_{1}(x),\quad 
    \hat\psi_2(x) =\psi_{2}(x,y)-2y\partial_{x}{\hat\psi}_1(x)
\end{align}
Then we can rewrite the modular Hamiltonian as \begin{align}\label{Hhat}
    &{\hat H}=\int dx dx'\hat\psi_{a}(x) {\hat H}^{(ab)}(x,y;x',y')\hat\psi_{b}(x'), \quad a,b=1,2,  \\
   &{\hat H}^{(22)}=0,\quad {\hat H}^{(12)}={\hat H}^{(21)}= H^{(12)}  \label{Hstructure1}=H^{(21)},\quad {\hat H}^{(11)}=H^{(11)}-4y \partial_xH^{(21)} .
\end{align}

To compute the entanglement entropy, we can discretize the direction $x$ so that \eqref{Hstructure} takes the following form
\begin{equation}
\mathcal{H}=\sum_{m,n\in\mathbb{Z}}\psi_{a,m}(y)H_{mn}^{(ab)}(y,y')\psi_{b,n}(y'),\quad H_{mn}^{(22)}=0\label{modularh00}
\end{equation}
where $m$ labels the discrete coordinate $x_{m}$. 
Then the reality, self-adjoint, and antisymmetric conditions become the following relations \begin{align}
    H^{(ab)*}_{nm}(y',y)=H^{(ba)}_{mn}(y,y')=H^{(ab)}_{mn}(y,y')=-H^{(ab)}_{nm}(y',y).
\end{align}

\subsection{Coherent states and Grassmann integral}
\label{subsec: coherent}
In this subsection, we use the coherent state method to compute the entanglement entropy for multiple intervals. 
Let us first consider a pair of 2D free fermions on a lattice, with the following anti-commutation relations
\begin{equation}\label{antichi}
    \{\hat \gamma_m,\hat \beta_n\}=\delta_{mn},\ \ \{\hat \beta_m,\hat \beta_n\}=0,\ \ \ \{\hat \gamma_m,\hat \gamma_n\}=0,
\end{equation}
where $\hat \beta_n$ denotes the fermion  on the $n$-th site. Note that the subsequent discussion only relies on the above commutation relations and thus can be applied to any free fermion model. In particular, the conclusion applies for both the 2D chiral fermion model and the 
BMS free fermion model. Coherent states are defined as
\begin{equation}\label{codef}
|\eta\rangle\equiv e^{-\sum\eta_n\hat \gamma_n}|\Omega\rangle,\quad \langle\bar\eta|=\langle\Omega|e^{-\sum\hat \beta_n\bar\eta_n}, \quad  \hat \beta_n|\Omega\rangle=0 
\end{equation}
which are the eigenstates of $\hat \beta_n$ with Grassmannian valued eigenvalue $\eta_n$,
\begin{equation}
   \hat \beta_n |\eta\rangle=\eta_n|\eta\rangle.
\end{equation}
The inner product between two coherent states is given by 
\begin{equation}
    \langle \bar \eta|\eta\rangle=e^{-\bar\eta^T\eta}\equiv e^{-\sum_i\bar\eta_n\eta_n}
\end{equation}
where $\bar\eta$ and $\eta$ are understood as independent Grassmannian numbers throughout this paper.
In this basis, the trace of an operator $\mathcal O$ is
\begin{equation}
    \tr \mathcal O=\int D\bar\eta D\eta\, e^{-\bar\eta^T\eta}\langle -\bar\eta|\mathcal O|\eta\rangle
\end{equation}
In particular, the operator 
\begin{equation}
   e^{-\hat \gamma^T H\hat \beta}\equiv  e^{-\sum H_{mn}\hat \gamma_m\hat \beta_n}=1+\hat \gamma^T(e^{-H}-1)\hat \beta
\end{equation}
has trace 
\begin{equation}\label{H12int}
    {\text tr} e^{-\hat \gamma^T H\hat \beta}=\int D\bar\eta D\eta\, e^{-\bar\eta^T (1+e^{-H})\eta} =\det (1+e^{-H}).
\end{equation}
The above result can be used to compute the modular Hamiltonian and its relation to the correlation functions. 
As we have argued in section 4.1, the modular Hamiltonian can be written as 
\begin{align}
    \hat H=\hat \gamma^T H\hat \beta+\hat \beta^T H^{(11)} \hat \beta 
\end{align}
where  $\hat \beta={\hat\psi_1\over \sqrt{2\pi}}, \hat \gamma={\hat\psi_2\over\sqrt{2\pi}}.$
Using the anticommutative property of Grassmannian variables, we learn that the $H^{(11)}$ term does not contribute to the path integral. Further using \eqref{H12int}, we obtain the un-renormalized partition function 
\begin{align}
    Z\equiv \tr e^{-\hat H}=\det (1+e^{-H})
\end{align}
so that the reduced density matrix is given by 
\begin{equation}
    \rho_{\mathcal A}=\frac{e^{-\hat \gamma^T H\hat \beta-\hat \beta^T H^{(11)}\hat \beta}}{\tr e^{-\hat \gamma^T H\hat \beta-\hat \beta^T H^{(11)}\hat \beta}}.
\end{equation}
The Wightman function of two fermions can also be calculated in this basis, and the result is given by 
\begin{equation}
{C}_{mn}\equiv\langle\hat \gamma_m\hat \beta_n\rangle=\frac{\tr (e^{-\hat H}\hat \gamma_m\hat \beta_n)}{Z}=(1+e^{-H})^{-1}_{mn}.
\end{equation}
The above result establishes a matrix relation between the two point function and the modular Hamiltonian,
\begin{equation}\label{CH}
    (1+e^{-H})^{-1}=C,\ \ H=-\log(C^{-1}-1)
\end{equation}
which leads to the entanglement entropy
\begin{align}\label{SEEC}
    S&=-\tr\rho_{\mathcal A}\log\rho_{\mathcal A}=\tr (CH)-\log\det (1+e^{-H})\nonumber\\
    &=-\tr[(1-C)\log(1-C)+C\log C].
\end{align}
Note that the trace here is taken over the matrix indices. From the above calculation, we learn that the term $H^{(11)}$ does not affect the calculation of the entanglement entropy, which is determined totally by $H=2H^{(21)}$ and then by $C$.
In Appendix \ref{appendixA}, we use the method of diagonalization and obtain the same relation \eqref{CH} and entanglement entropy \eqref{SEEC}.
Note that the relation between the Wightman function and the entanglement entropy is the same as that in the free fermion CFT$_2$ \cite{Casini:2009vk}. 
The explicit result of the entanglement entropy will depend on the theory and also the state, which we shall discuss momentarily.

\subsection{Explicit results}
The relation  \eqref{SEEC} enables us to compute the entanglement entropy using the two-point correlators. Note that the correlation function depends on the choice of the vacuum. 
In the following, we will explicitly carry out the computation in the following two choices of the vacuum.

\paragraph{Induced vacuum}
We first consider the induced vacuum under which the correlator reads \cite{Hao:2022xhq}
\begin{equation}
C_{mn}=\delta_{mn}
\end{equation}
due to its ultra-local nature. Consider an interval $\mathcal A$ whose $x$ coordinates have range $V=\cup_i [a_i,b_i]$, the two-point function restricted in this region is given by 
\begin{align}
    C_{mn}=\Big\{\,\begin{aligned}& \delta_{mn},\quad \text{for} \  x_m,x_n\in V;\\
     &0,\quad \text{otherwise}
    \end{aligned}
\end{align}
which is already diagonal with eigenvalues $0$ or $1$. 
Plugging the above result into \eqref{SEEC}, we find that the entanglement entropy
\begin{equation}
    S=0
\end{equation}
which indicates that it is a pure state for an arbitrary interval. This is consistent with the fact that BMSFT on the induced vacuum is the product state pointwise.

\subsubsection*{The highest weight vacuum}
The calculation of entanglement entropy in the highest weight vacuum is similar to the relativistic free fermion model \cite{Casini:2009sr}.  
In the discrete case, the two-point functions on the lattice can be calculated numerically and then restricted to the desired interval. Then the eigenvalues of this smaller matrix can also be numerically computed to obtain the entanglement entropy. In the continuous case, the resolvent $R$ is introduced to determine the spectrum of the two-point function operator and to express the entanglement entropy in a convenient form:
\begin{equation}
    R(\lambda)=\left(C-\frac{1}{2}+\lambda\right)^{-1}
\end{equation}
where $\lambda$ is an auxiliary parameter.
Then the entanglement entropy can be expressed in terms of $R$ as
\begin{equation}\label{SEE}
    S=-\int_{1/2}^{\infty}d\lambda \,{\text{tr}}\left[(\lambda-1/2)(R(\lambda)-R(-\lambda))-\frac{2\lambda}{\lambda+1/2}\right]
\end{equation}
where the trace is taken over the spectrum of $R$. The resolvent depends on both the specific expression of the two-point function and the region of interest.

Now let us consider the global vacuum state on a multi-interval region $\mathcal{A}$.  
In the highest vacuum, the Wightman function restricted on $\mathcal{A}$ can be obtained from the $x$-ordered two point function \eqref{2pfh},
\begin{equation}
    C=\frac{1}{2}\delta(x-x')-\frac{i}{2\pi}\frac{1}{x-x'},\quad  x, x'\in V=\cup_i^n V_i,  \quad V_i=[a_i,b_j]
\end{equation}
which is again similar to the relativistic case \cite{Casini:2009sr}.
The resolvent on the $n$ disjoint interval $V$ can be obtained from the singular integral equations of the Cauchy type  \cite{Casini:2009vk,mikhlin1947ni,estrada2012singular}, and can be expressed as, 
\begin{align}
    R(\lambda)=\left(\lambda^2-\frac{1}{4}\right)^{-1}\left(\lambda\delta(x-x')+\frac{i}{2\pi( x-x')}e^{-\frac{i}{2\pi} \left(\log\frac{\lambda-1/2}{\lambda+1/2}\right)(\log z-\log z')}\right)
\end{align}
where $z=z(x),z'=z(x')$ is determined by   
 the uniformazation map \eqref{un} which we reproduce here for convenience  \begin{align} z(x)\equiv -{\prod^n_{i=1}(x-a_i)\over \prod^n_{i=1}(x-b_i)}\,.\end{align} Plugging  the above resolvent into \eqref{SEE}, we find that the terms proportional to the delta function cancel out, and the entanglement entropy is given by,
\begin{align}
    S=&\int_V dx\lim_{x'\rightarrow x}\int_{1/2}^{\infty} d\lambda\frac{\sin(\frac{1}{2\pi}(\log\frac{\lambda-1/2}{\lambda+1/2})(\log z-\log z'))}{-\pi(\lambda+1/2)(x-x')}\nonumber\\
    =&\frac{1}{6}\Big(\sum_{i,j}\log|b_i-a_j|-\sum_{i<j}\log|a_i-a_j|-\sum_{i<j}\log|b_i-b_j|-n\log\epsilon\Big)\label{Seefn}
\end{align}
where $\epsilon$ is the cutoff around the endpoints of the intervals.

In the highest weight vacuum, the entanglement entropy for a single interval in the free BMS fermion theory is given by \eqref{Seefn} with $n=1$, which is indeed consistent with the general result \eqref{singleEE} obtained from the replica trick or the Rindler method \cite{Bagchi:2014iea,Jiang:2017ecm} with central charges $c_L=1,\,c_M=0$. 
In this case, the entanglement entropy only depends on the separation in $x$, and takes the form of a chiral CFT. 
In fact, the result with $n$-intervals \eqref{Seefn} also agrees with that of a chiral CFT. The reason is that the free BMS fermion theory has central charge $c_M=0$, and the contribution of the separation in $y$ is proportional to $c_M$, as can be seen explicitly from \eqref{singleEE}. 

\subsection{Comparison to the swing surface proposal}  

It has been proposed in \cite{Jiang:2017ecm,Apolo:2020bld} that entanglement entropy in BMSFT can be calculated holographically by minimizing a functional of the swing surface. For Einstein gravity, we have $c_L=0,\,c_M\neq0$, and the holographic entanglement entropy is proportional to the area of the swing surface. By adding a topologically Chern-Simon term to Einstein gravity, we obtain the so-called topologically massive gravity \cite{deser1982three}, which has $c_L\neq0$. In the limiting case of flat chiral gravity \cite{Bagchi:2012yk} with only topologically Chern-Simons term, we have $c_L\neq0,\,c_M=0$, which is qualitatively similar to the free BMS fermion theory. 

We have shown that our explicit result of entanglement entropy on a single interval in the highest weight vacuum agrees with the Rindler method result \eqref{singleEE}. The later of which was shown to be consistent with the swing surface proposal \cite{Jiang:2017ecm,Apolo:2020bld}. 

Now let us consider two disjoint intervals. 
The entanglement entropy in free BMS fermion for two disjoint intervals in the highest weight vacuum can be obtained from the general result \eqref{Seefn} by taking $n=2$,
\begin{align}
    S
=\frac{1}{6}\log{4ab(a-b)^2\over(a+b)^2\epsilon^2},
\end{align}
where we have chosen the parameters consistent with \eqref{xends}
\begin{equation}
    a_1=-b,\ \ a_2=a,\ \ b_1=-a,\ \ b_2=b.
\end{equation}
In this case, the mutual information reads
\begin{align}\label{MI}
    I=S-S_{[-b,-a]}-S_{[a,b]}={1\over6}\log{4ab\over(a+b)^2}=\frac{1}{6}\log(1-x),
\end{align}
where $x$ is the cross ratio
\begin{equation}
    x=\frac{x_{12}x_{34}}{x_{13}x_{24}}={(b-a)^2\over(a+b)^2}.
\end{equation}
When the intervals are far apart from each other $x\to 0$, the mutual information approaches zero $I\to0$. When the intervals are close to each other as $x\to1$, the mutual information gets divergent due to the UV behavior. 

Holographically, the entanglement entropy for the interval $\mathcal A$ is the minimum value between two possible ways of connecting the four end points in the bulk by the swing surface. Using the holographic result of single intervals with $c_M=0$ in \cite{Jiang:2017ecm}, we have
\begin{align}
    S&=\min\big[\,S_{[-b,-a]}+S_{[a,b]},\quad S_{[-b,b]}+S_{[-a,a]} \,  \big]\nonumber\\[.5ex]
    &=\left\{\begin{aligned}
   &{c_L\over6}\log (b-a)^2,\quad x={(b-a)^2\over (b+a)^2}\le{1\over2},\\[.5ex]
    &{c_L\over6}\log (4ab),\quad\quad\, x>{1\over2}.
    \end{aligned}\right.
\end{align}
A phase transition appears at $(b-a)^2=4ab$ in which case $x={1\over2}$. 
Then the mutual information is given by 
\begin{equation}\label{hMI}
    I=\left\{\begin{aligned}
        &0,\quad\quad\quad\quad\quad\quad\, x\le{1\over2},\\[.5ex]
        &{c_L\over6}\log {1-x\over x},\quad\ x>{1\over2}.
    \end{aligned}\right.
\end{equation}

When comparing the free BMS fermion result \eqref{MI} and the holographic result \eqref{hMI}, we observe that both of them vanish as $x\to0$ and share the same divergence as $x\to1$. It is important to note that the free fermion result \eqref{MI} is quantum, whereas the holographic calculation \eqref{hMI} is classical and, therefore, may not fully reproduce the quantum answer.

\acknowledgments
We are grateful to Liangyu Chen, Kristan Jensen, Shanming Ruan, Joan Si\'mon, Boyang Yu and Yuan Zhong for helpful discussions. 
The work is supported by the NSFC special fund for theoretical physics No.\,12447108, the NSFC general program No.\,12175238, and the national key research and development program of China No.\,2020YFA0713000.

\appendix

\section{Entanglement entropy from diagonalization}\label{appendixA}
In the following, we use the method of diagonalization to work out the relation between entanglement entropy from correlation functions. 
To do so, we first consider the single-interval case where the $H^{(11)}$ term can be removed with the Rindler transformation, and we follow the discussion in \cite{Casini:2009sr}.
As we discussed in section 3, local modular Hamiltonians can all be obtained from the half-line result in the $(\hat z, \hat w)$ plane \eqref{Hzhat}. For convenience, we introduce the Rindler coordinate 
\begin{equation}
    \hat z=e^{2\pi\sigma}
\end{equation}
so that the modular Hamiltonian becomes
\begin{equation}\label{HhatApp}
\hat{\mathcal{H}}
={i\over2}\int_{-\infty}^\infty d\sigma{\hat\gamma}H{\hat\beta},  
\end{equation}
where we have rescaled the spinors by 
\begin{align}
    \hat\beta={\hat\psi_{1}\over\sqrt{2\pi}}, \quad \hat\gamma={\hat\psi_{2}\over\sqrt{2\pi}} 
\end{align}
so that the anti-commutation relation  \eqref{anticm} between the fermionic fields is equivalent to \begin{align}\label{rescaled}
\{\hat\beta(\sigma_{1}),\hat\gamma(\sigma_{2})\}=\delta(\sigma_{1}-\sigma_{2}), \quad \{\hat\beta(\sigma_{1}),\hat\beta(\sigma_{2})\}=\{\hat\gamma(\sigma_{1}),\hat\gamma(\sigma_{2})\}=0. \end{align}
After discretizing the coordinate $\sigma\to\sigma_m$, 
we have the following Bogoliubov transformation
\begin{align}\label{betamodes}
\hat{\beta}_{m} & =\sum_{k}\varphi_{m,k}b_{k},\quad
\hat{\gamma}_{m}
=\sum_{k}\varphi_{m,k}a_{k}
\end{align}
where $ \varphi_{m,k}$ forms an orthonormal basis that diagonalizes the modular Hamiltonian, so that the kernel can be written in the matrix form as 
\begin{align}
\hat{\mathcal{H}}=\sum_{m,n}\hat{\gamma}_mH_{mn}{\hat\beta}_{n}
=\hat \gamma^T H \hat\beta , \quad H_{mn} & =\sum_{k}\varphi_{m,k}\varphi_{n,k}^{*}\varepsilon_{k}\label{kernel}
\end{align}
which acts on the modes as
\begin{align}
H_{mn}\varphi_{n,k}={}&\varepsilon_{k}\varphi_{m,k}.
\end{align}
Using the discrete version of the canonical commutation relation $\{\hat\beta_{m},\hat\gamma_{n}\}=\delta_{m,n}$, we find that the anti-commutation relations between the coefficients take the standard form, with the non-vanishing commutators:
\begin{align}
\left\{ a_k,b_{l}\right\} &=\delta_{k+l,0},\quad \left\{ a_k,a_{l}\right\} =0, \quad \left\{ b_k,b_{l}\right\} =0.
\end{align}
Then the modular Hamiltonian \eqref{HhatApp} can be diagonalized,
\begin{equation}
\hat{\mathcal{H}}
=\sum_{k}\varepsilon_ka_{-k}b_k,
\end{equation}
where we choose $b_k$s as the annihilation operators to define the vacuum.
The normalized reduced density matrix with 
$\tr\rho=1$ is then given by 
\begin{align}
\rho_{\mathcal A}&=\hat Z^{-1} e^{-\hat {\mathcal H}}=\prod_{k} {e^{-\varepsilon_ka_{-k}b_k}\over1+e^{-\varepsilon_k}},\quad 
   {\hat Z}=\det(1+e^{-H}).
\end{align}
Insert the above density matrix and the modes expansion for fermion fields \eqref{betamodes} into the trace expression of the Wightman functions of the free BMS fermion system, we find
\begin{align} &\tr\left(\rho_{\mathcal A}{\hat\beta}_{m}{\hat\beta}_{n}\right)=0,\quad \tr\left(\rho_{\mathcal A}\hat{\gamma}_{m}\hat{\gamma}_{n}\right)=0,\\
   & C_{mn}\equiv\langle 0|\hat{\gamma}_{m}\hat{\beta}_{n}|0\rangle   =\tr\left(\rho_{\mathcal A}\hat{\gamma}_{m}\hat{\beta}_{n}\right)=\sum_{k}\varphi_{m,k}^{*}\varphi_{n,k}\frac{1}{e^{\varepsilon_{k}}+1},\label{2-pt}
\end{align}
where $|0\rangle$ denotes the vacuum state of the theory, and the trace in the second equality is over the Hilbert space in region $\mathcal A$. 
We find that the relation between the eigenvalues $\varepsilon_k$ of the kernal $H$ and the eigenvalues $\zeta_{k}$ of the 2-point function $C$ 
\begin{align}
\zeta_{k} & =\frac{1}{e^{\varepsilon_{k}}+1},\ \ \varepsilon_{k} =\ln\left(\zeta_{k}^{-1}-1\right),
\end{align}
which indicates a matrix condition
\begin{equation}
H=-H^{T}=-\ln\left[\left(C\right)^{-1}-1\right]\label{HC}.
\end{equation}
Then the entanglement entropy is given by  
\begin{align}\label{C2SEE}
    S&=-\tr\rho_{\mathcal A}\log\rho _{\mathcal A}=\sum_{k} \left(\log(1+e^{-\varepsilon_k})+\frac{\varepsilon_{k}e^{-\varepsilon_{k}}}{1+e^{-\varepsilon_k }}\right)\nonumber\\
    &=-\tr[(1-C)\log(1-C)+C\log C],
\end{align}
where $C$ is the Wightman function and the trace is over states in the subregion.
The above formula establishes a relationship between the entanglement entropy and the two-point functions in the BMS free fermion model. 
This provides an alternative derivation of the result in subsection \ref{subsec: coherent}. 

\bibliographystyle{JHEP.bst}
\bibliography{refs.bib}

\end{document}